\documentclass[fleqn,usenatbib]{mnras}
\usepackage{mathptmx}

\usepackage[T1]{fontenc}
\usepackage{ae,aecompl}

\usepackage{graphicx}	
\usepackage{amsmath}	
\usepackage{amssymb}	
\usepackage[flushleft]{threeparttable}
\usepackage{hyperref}
\hypersetup{breaklinks=true, colorlinks=true, urlcolor={blue}, citecolor={blue}, linkcolor={blue}}
\usepackage{afterpage}

\newcommand{\chandra}{\emph{Chandra}}

\newcommand{\xmm}{\emph{XMM-Newton}}

\newcommand{\lumcgs}{erg~s$^{-1}$}

\title[Pulse Profile of Magnetars]{A systematic study of soft X-ray pulse profiles of magnetars in quiescence}

\author[C.-P. Hu et al.]{
Chin-Ping Hu$^{1,2}$\thanks{E-mail: cphu@kusastro.kyoto-u.ac.jp}\thanks{JSPS International Research Fellow}
C.-Y. Ng$^{1}$,
and Wynn C. G. Ho$^{3,4}$
\\
$^{1}$Department of Physics, The University of Hong Kong, Pokfulam Road, Hong Kong\\
$^{2}$Department of Astronomy, Kyoto University, Oiwake-cho, Sakyo-ku, Kyoto 606-8502, Japan\\
$^{3}$Department of Physics and Astronomy, Haverford College, 370 Lancaster Avenue, Haverford, PA 19041, USA\\
$^{4}$Mathematical Sciences, Physics and Astronomy, and STAG Research Centre, University of Southampton, Southampton SO17 1BJ, UK\\
}

\date{Accepted 2019 February 18. Received 2019 February 7; in original form 2018 November 13}

\pubyear{2019}

\begin{document}
\label{firstpage}
\pagerange{\pageref{firstpage}--\pageref{lastpage}}
\maketitle

\begin{abstract}
Magnetars are neutron stars with extremely high surface magnetic fields. They show diverse X-ray pulse profiles in the quiescent state. We perform a systematic Fourier analysis of their soft X-ray pulse profiles. We find that most magnetars have a single-peaked profile and hence have low amplitudes of the second Fourier harmonics ($A_2$). On the other hand, the pulsed fraction (PF) spreads over a wide range. We compared the results with theoretical profiles assuming various surface hotspot asymmetries, viewing geometries, and beaming functions. We found that a single value of the intensity ratio $r$ between two antipodal hotspots is unable to reproduce the observed distribution of $A_2$ and PF for all magnetars. The inferred $r$ is probably anticorrelated with the thermal luminosity, implying that high-luminosity magnetars tend to have two symmetric hotspots. Our results are consistent with theoretical predictions, for which the existence of an evolving toroidal magnetic field breaks the symmetry of the surface temperature.

\end{abstract}
\begin{keywords}
stars: magnetars -- stars: neutron -- pulsars: general -- X-rays: stars
\end{keywords}



\section{Introduction}
Magnetars are isolated neutron stars (NSs) exhibiting dramatic timing variabilities and are believed to have extraordinarily high surface magnetic fields \citep[see review by][]{KaspiB2017}. They have spin periods clustering in the range of $P=2$--$12$\,s, large spin-down-inferred dipolar magnetic fields of $B=10^{12}$--$10^{16}$\,G, and high surface luminosities of $L\approx 10^{31}$--$10^{36}$\lumcgs\ in the quiescent state. The most remarkable feature of magnetars is burst in soft gamma-ray/hard X-ray bands. The intense bursting epochs are usually accompanied with outbursts, during which the persistent X-ray luminosity increases dramatically on a short timescale and then decays slowly for months to years. The mechanisms triggering the outburst remain a puzzle. The energy could be injected from the deep layer of the crust through magnetic dissipation or from the interaction between the currents along the twisted magnetic field lines and the stellar surface \citep{LyubarskyET2002, BeloborodovT2007, Beloborodov2009, PonsR2012}. The bombardment of particles from the magnetosphere produces additional hotspots and provides a possible heating source \citep{BeloborodovT2007, BeloborodovL2016}. Observations show that the hotspot temperature decreases and the corresponding area shrinks during the flux relaxation of the outburst \citep[see e.g.,][]{ReaIP2013, ZelatiRP2015, MongN2017}. The emerging hotspots generally cause extra peaks in the pulse profile, while the strength of the peak decreases as the flux decreases \citep[see e.g., ][]{RodriguezIE2014}. The shape of the pulse profile usually evolves back to the pre-outburst state within months to years. Several magnetars, such as 4U 0142$+$61, have no recorded major outburst, although they could have subtle flux variability. Their quiescent fluxes and profiles show moderate variabilities but are relatively stable compared with the change during outbursts \citep{DibKG2007}.

The high X-ray luminosity of magnetars is believed to be powered by the decay of the strong $B$ field, although the conversion mechanism is elusive. The magneto-thermal evolution model suggests that magnetic energy is transferred to heat through the dissipation process in the crust. This model explains the systematically high surface temperature of magnetars and unifies the temperature evolution of various NS populations \citep{KaminkerYP2006, KaminkerPY2009, PonsMG2009, PernaP2011, HoGA2012, ViganoRP2013, KaminkerKP2014}. Instead of having a pure dipole $B$ field, magnetars are believed to have complex $B$-field structures such as a toroidal component \citep{ThompsonLK2002, PavanTZ2009, ViganoRP2013}. These components make the total $B$ field stronger than the observable dipolar term and enhance the decay of the $B$-field strength. This process increases the heating of the NS surface and extends the cooling time-scale. It also breaks the $B$-field symmetry, resulting in an asymmetric surface temperature distribution owing to either the suppression of the temperature of one hotspot or the migration of the coldest/hottest part from the magnetic equator/poles \citep{NgKH2012, ViganoRP2013}. The change in the surface temperature symmetry can be inferred from the quiescent thermal pulse profile. Previous studies revealed highly modulated single-peaked pulse profiles for some magnetars \citep[e.g., ][]{TamGD2008, ZhouCL2014}. It could indicate an asymmetric surface temperature distribution and cannot be explained by two hotspots located at the magnetic poles with similar luminosities \citep{PernaVP2013}. For a magnetar with a strong initial toroidal field, it is expected that the pulse profile is time-dependent and could be correlated with the $B$-field strength, age, or the thermal luminosity. This motivated us to perform a systematic analysis of magnetar thermal pulse profiles. Similar analyses have been applied to several individual magnetars to trace the pulse-profile evolution on various time-scales \citep[e.g.,][]{DedeoPN2001, TamGD2008, SasmazG2013}, but a comprehensive investigation of all currently known magnetars is lacking. We parameterize the quiescent soft X-ray pulse profiles of magnetars and investigate their connection with physical parameters.

We introduce the source selection and the basic reduction of the \chandra\ and \xmm\ data in Section \ref{data_reduction}. Then, we describe the pulse-profile analysis method in Section \ref{method}. The results for individual sources and the observed distribution of pulse-profile parameters are described in Section \ref{result}. We present simulations and discuss the evolution of the surface temperature anisotropy in Section \ref{simulation}. The connection between the observed pulse profiles and the surface intensity distribution is described in Section \ref{discussion}. Finally, we summarize this work in Section \ref{summary}. We also found that the previously reported timing solutions of CXOU J171405.7$-$381031 are highly variable. We performed a timing analysis on several new datasets to update the $\dot{P}$ value in Appendix \ref{new_pdot}.

\begin{figure}
\includegraphics[width=0.47\textwidth]{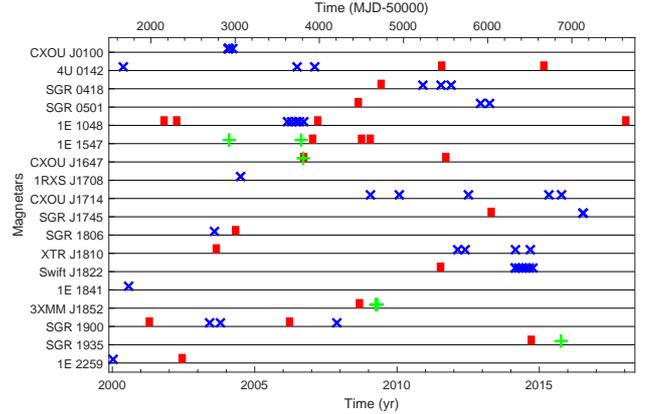}
\caption{Outburst history of all magnetars and the distribution of data sets used in this analysis. The red squares denote the onset epochs of the outbursts, the blue crosses are the \chandra\ observation epochs, and the green plusses are the \xmm\ observation epochs for individual magnetars. Only those data sets used in this analysis are plotted. \label{observation_log}}
\end{figure}

\begin{table*}
	\centering
	\caption{Data sets used in this analysis.}
	\label{summary_dataset}
	\begin{threeparttable}
	\begin{tabular}{lcl} 
		\hline
		Name & Instrument/Mode$^a$ & Dataset (ObsID)\\
		\hline
CXOU~J010043.1$-$721134 & \chandra/TE & \href{http://cda.harvard.edu/chaser/viewerContents.do?obsid=1881}{1881}, \href{http://cda.harvard.edu/chaser/viewerContents.do?obsid=4616}{4616}, \href{http://cda.harvard.edu/chaser/viewerContents.do?obsid=4617}{4617}, \href{http://cda.harvard.edu/chaser/viewerContents.do?obsid=4618}{4618}, \href{http://cda.harvard.edu/chaser/viewerContents.do?obsid=4619}{4619}, \href{http://cda.harvard.edu/chaser/viewerContents.do?obsid=4620}{4620} \\
4U~0142+61 & \chandra/CC & \href{http://cda.harvard.edu/chaser/viewerContents.do?obsid=724}{724}, \href{http://cda.harvard.edu/chaser/viewerContents.do?obsid=6723}{6723}, \href{http://cda.harvard.edu/chaser/viewerContents.do?obsid=7659}{7659} \\
SGR~0418+5729 & \chandra/TE & \href{http://cda.harvard.edu/chaser/viewerContents.do?obsid=13148}{13148}, \href{http://cda.harvard.edu/chaser/viewerContents.do?obsid=13235}{13235}, \href{http://cda.harvard.edu/chaser/viewerContents.do?obsid=13236}{13236} \\
SGR~0501+4516 & \chandra/TE & \href{http://cda.harvard.edu/chaser/viewerContents.do?obsid=14811}{14811},  \href{http://cda.harvard.edu/chaser/viewerContents.do?obsid=15564}{15564} \\
1E~1048.1$-$5937 & \chandra/CC & \href{http://cda.harvard.edu/chaser/viewerContents.do?obsid=6733}{6733}, \href{http://cda.harvard.edu/chaser/viewerContents.do?obsid=6734}{6734}, \href{http://cda.harvard.edu/chaser/viewerContents.do?obsid=6735}{6735}, \href{http://cda.harvard.edu/chaser/viewerContents.do?obsid=6736}{6736}, \href{http://cda.harvard.edu/chaser/viewerContents.do?obsid=7347}{7347} \\
1E~1547.0$-$5408 & \emph{XMM} PN/FF & \href{http://nxsa.esac.esa.int/nxsa-web/\#obsid=0604880101}{0604880101} \\
CXOU~J164710.2$-$455216 & \emph{XMM} PN/FF & \href{http://nxsa.esac.esa.int/nxsa-web/\#obsid=0404340101}{0404340101}\\
1RXS~J170849.0$-$400910 & \chandra/CC & \href{http://cda.harvard.edu/chaser/viewerContents.do?obsid=4605}{4605} \\
CXOU~J171405.7$-$381031 & \chandra/CC & \href{http://cda.harvard.edu/chaser/viewerContents.do?obsid=10113}{10113}, \href{http://cda.harvard.edu/chaser/viewerContents.do?obsid=11233}{11233}, \href{http://cda.harvard.edu/chaser/viewerContents.do?obsid=13749}{13749}, \href{http://cda.harvard.edu/chaser/viewerContents.do?obsid=16762}{16762}, \href{http://cda.harvard.edu/chaser/viewerContents.do?obsid=16763}{16763} \\
SGR~J1745$-$2900$^b$& \chandra/TE & \href{http://cda.harvard.edu/chaser/viewerContents.do?obsid=18731}{18731}, \href{http://cda.harvard.edu/chaser/viewerContents.do?obsid=18732}{18732} \\
SGR~1806$-$20$^b$ & \chandra/CC & \href{http://cda.harvard.edu/chaser/viewerContents.do?obsid=4443}{4443}  \\
XTE~J1810$-$197 & \chandra/TE & \href{http://cda.harvard.edu/chaser/viewerContents.do?obsid=13746}{13746}, \href{http://cda.harvard.edu/chaser/viewerContents.do?obsid=13747}{13747}, \href{http://cda.harvard.edu/chaser/viewerContents.do?obsid=15870}{15870}, \href{http://cda.harvard.edu/chaser/viewerContents.do?obsid=15871}{15871}  \\
Swift~J1822.3$-$1606 & \chandra/TE & \href{http://cda.harvard.edu/chaser/viewerContents.do?obsid=14819}{14819}, \href{http://cda.harvard.edu/chaser/viewerContents.do?obsid=15988}{15988}, \href{http://cda.harvard.edu/chaser/viewerContents.do?obsid=15989}{15989}, \href{http://cda.harvard.edu/chaser/viewerContents.do?obsid=15990}{15990}, \href{http://cda.harvard.edu/chaser/viewerContents.do?obsid=15991}{15991}, \href{http://cda.harvard.edu/chaser/viewerContents.do?obsid=15992}{15992}, \href{http://cda.harvard.edu/chaser/viewerContents.do?obsid=15993}{15993} \\
1E~1841$-$045 & \chandra/CC & \href{http://cda.harvard.edu/chaser/viewerContents.do?obsid=730}{730} \\ 
3XMM~J185246.6+003317 & \emph{XMM} MOS/FF & \href{http://nxsa.esac.esa.int/nxsa-web/\#obsid=0550671301}{0550671301}, \href{http://nxsa.esac.esa.int/nxsa-web/\#obsid=0550671801}{0550671801}, \href{http://nxsa.esac.esa.int/nxsa-web/\#obsid=0550671901}{0550671901} \\
SGR~1900+14 & \chandra/CC & \href{http://cda.harvard.edu/chaser/viewerContents.do?obsid=3863}{3863}, \href{http://cda.harvard.edu/chaser/viewerContents.do?obsid=3864}{3864}, \href{http://cda.harvard.edu/chaser/viewerContents.do?obsid=8215}{8215} \\
SGR~1935+2154 & \emph{XMM} PN/FF & \href{http://nxsa.esac.esa.int/nxsa-web/\#obsid=0764820201}{0764820201} \\
1E~2259+586 & \chandra/CC & \href{http://cda.harvard.edu/chaser/viewerContents.do?obsid=726}{726}\\
		\hline
	\end{tabular}
	\begin{tablenotes}
            \item[a] CC: \chandra\ ACIS continuous clocking mode, TE: \chandra\ ACIS timed-exposure $1/8$ subarray mode, FF: \xmm\ full-frame mode.
            \item[b] We extended the high energy boundary of the pulse profile to 4\,keV to boost the signal to noise ratio (see text).
        \end{tablenotes}
        \end{threeparttable}
\end{table*}

\section{Source Selection and X-ray Data Reduction}\label{data_reduction}
We selected the sample from the McGill OnlineMagnetar Catalog\footnote{\url{http://www.physics.mcgill.ca/~pulsar/magnetar/main.html}} \citep{OlausenK2014}. There are 23 confirmed magnetars to date. We utilized the data collected with \chandra\ and \xmm\ owing to their excellent sensitivity below 2\,keV. We checked the outburst history of magnetars from the Magnetar Outburst Online Catalog\footnote{\url{http://magnetars.ice.csic.es/}}\citep{CotiZelatiRP2018}.  We chose observations that were taken at least half a year after the onset of an outburst (or the first observation available since the outburst if the onset time is unknown) and have a flux that decreased to $\sim$10 per cent of the peak flux.  From a total of 23 magnetars, we analysed 18. We could not obtain pulse profiles for the remaining ones. For example, SGR~0526$-$66 showed a pulsating tail with $P\approx8$\,s after a giant flare \citep{ClineDP1980}. A similar period was marginally determined in two \chandra\ and one \xmm\ datasets with low significances \citep{Kulkarni2003, TiengoEM2009}. However, no stable pulse profile could be obtained from the datasets. Moreover, we did not detect any significant periodicity in PSR~J1622$-$4950, SGR~1627$-$41, and Swift~J1834.9$-$0846. Finally, SGR~1833$-$0832 was not detected in either \chandra\ and \xmm\ observations before its outburst. Fig.~\ref{observation_log} shows the outburst history of all sampled magnetars and the observation epoch of the datasets used in this research.

To investigate the shape of pulse profiles, observations with high timing resolution are necessary. Therefore, the data taken with the \chandra\ Advanced CCD Imaging Spectrometer (ACIS) in the continuous-clocking (CC) mode is preferred, because its timing resolution is as high as $2.85$\,ms.  For sources without CC mode observations, we utilized the $1/8$ subarray timed-exposure (TE) mode observations (timing resolution of $\sim0.4$\,s) in order to minimize the contamination from the surroundings. For the remaining sources without \chandra\ observations in quiescence, we investigated their pulse profiles with \xmm\ observations. The PN detector has a timing resolution of 74\,ms for the full-frame (FF) mode observation, while it is 2.6\,s for the MOS detectors.  Therefore, we mainly utilized the data obtained with the PN detector. We calculated the background-subtracted pulse profiles for all sampled magnetars. All the observations used in this analysis are summarized in Table \ref{summary_dataset}. 

We downloaded the \chandra\ data from the \chandra\ Data Archive\footnote{\url{http://cda.harvard.edu/chaser/}}, and reprocessed them using the pipeline `chandra\_repro' in the Chandra Interactive Analysis of Observations (CIAO) version 4.9 with the calibration data base (CALDB) version 4.7.3 \citep{FruscioneMA2006}. All the photon arrival times were corrected to the barycentre of the Solar system using the task `axbary' based on the JPL ephemeris DE405. We extracted the source photons from a 4-arcsec-wide box centred on the source for CC mode observations. The fractional encircled flux is energy-dependent, and this selection criterion contains an average of $\sim$90 per cent of the source flux. The source selection criteria used for the \chandra\ TE mode and \xmm\ observations also contain $\sim$90 per cent of the source flux. The background events were extracted from a box with the same size in a nearby region. For \chandra\ data taken in the $1/8$ sub-array TE mode, we extracted the source photons from a 2-arcsec-radius circular aperture centred on the source, and extracted background events from nearby source-free regions.

For \xmm, we downloaded the data from the \xmm\ Science Archive\footnote{\url{http://nxsa.esac.esa.int/nxsa-web/}} and reduced them using the \xmm\ Science Analysis Software (SAS) version 16.0.0. We reprocessed the PN data with current calibration files using `epproc'. After the reduction, we corrected the photon arrival time to the barycentre of the Solar system using `barycen' with the JPL ephemeris DE405. The time intervals with flaring particle background were filtered out. We then extracted the source photons from a circular aperture with a 40-arcsec radius, and the background events from nearby source-free regions. For 3XMM~J1852, all the observations were pointed to the nearby pulsar J1852+0040 and PN was operated in the small window mode. As a result, 3XMM~J1852 lied outside the PN field of view. We therefore utilized the MOS data. We used `emproc' to reprocess the data and performed all the necessary corrections to photon events as for the PN data. Then we combined the events collected from both the MOS1 and the MOS2 detector for the timing analysis. 

\begin{figure*}
\begin{center}
\begin{minipage}{0.32\linewidth}
\includegraphics[width=1.05\textwidth]{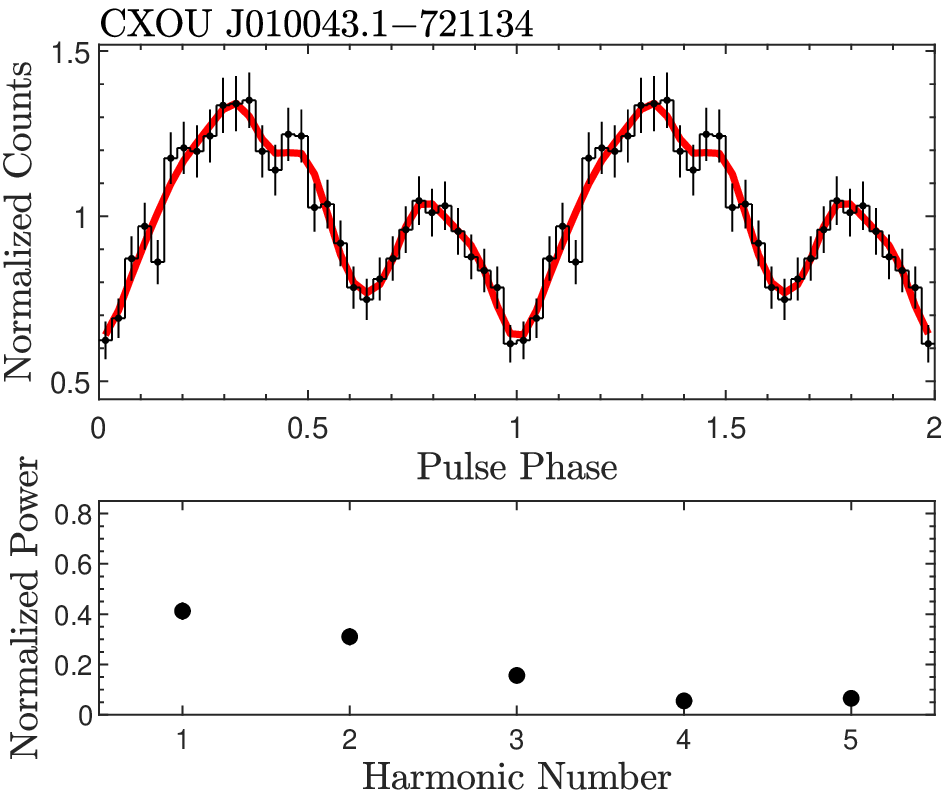}
\includegraphics[width=1.05\textwidth]{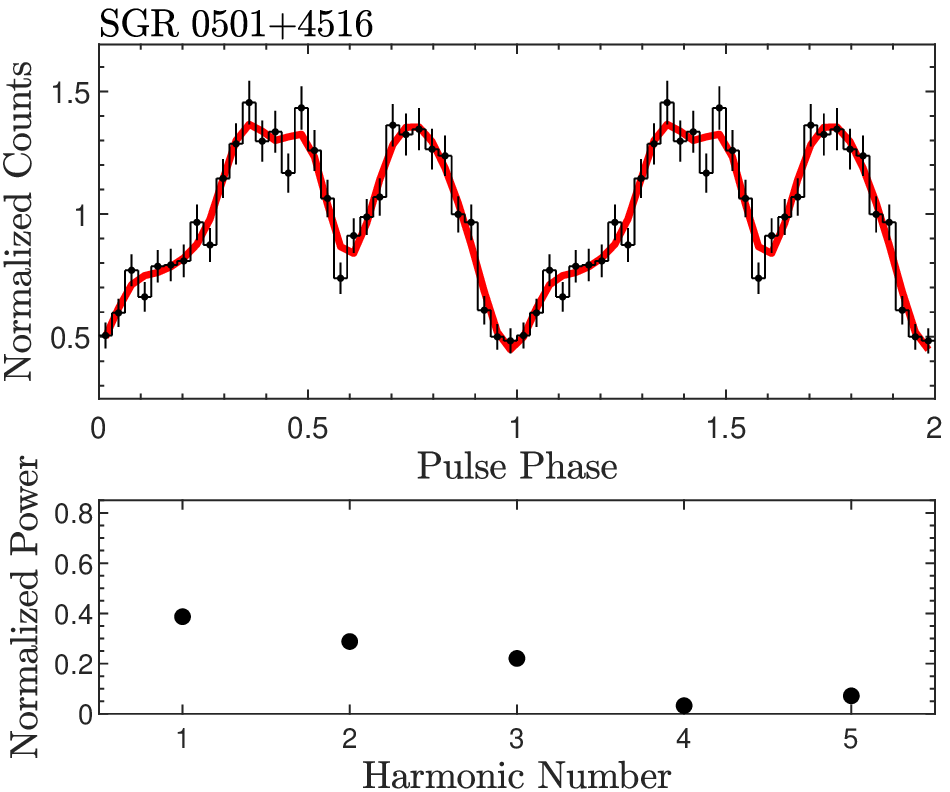}
\includegraphics[width=1.05\textwidth]{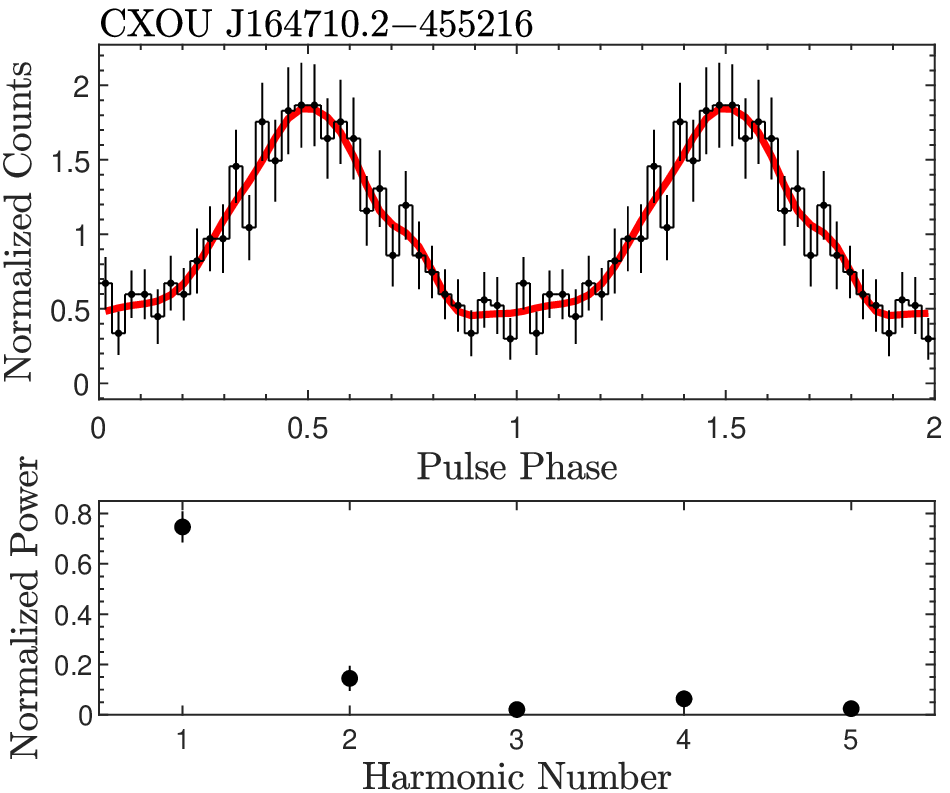}
\includegraphics[width=1.05\textwidth]{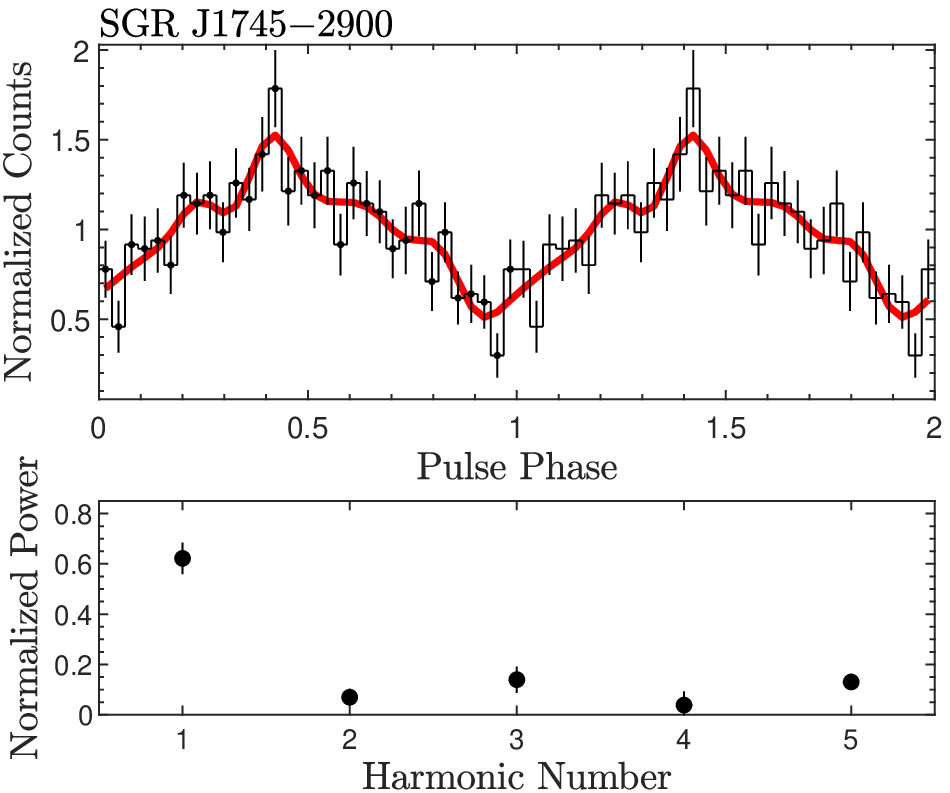}
\includegraphics[width=1.05\textwidth]{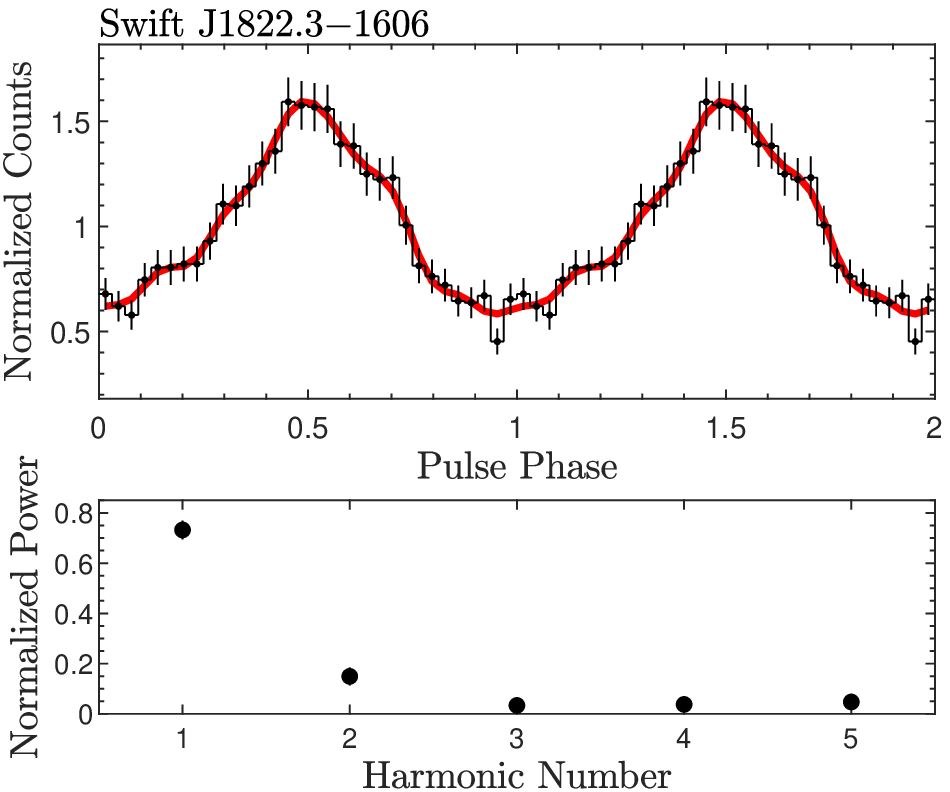}
\end{minipage}
\begin{minipage}{0.32\linewidth}
\includegraphics[width=1.05\textwidth]{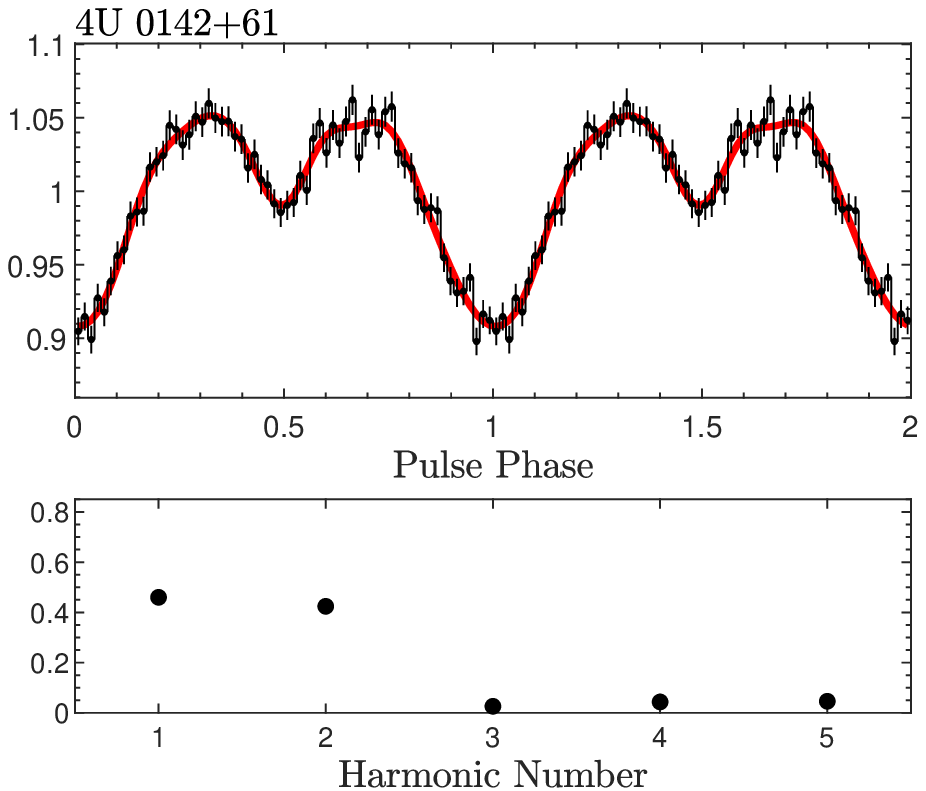}
\includegraphics[width=1.05\textwidth]{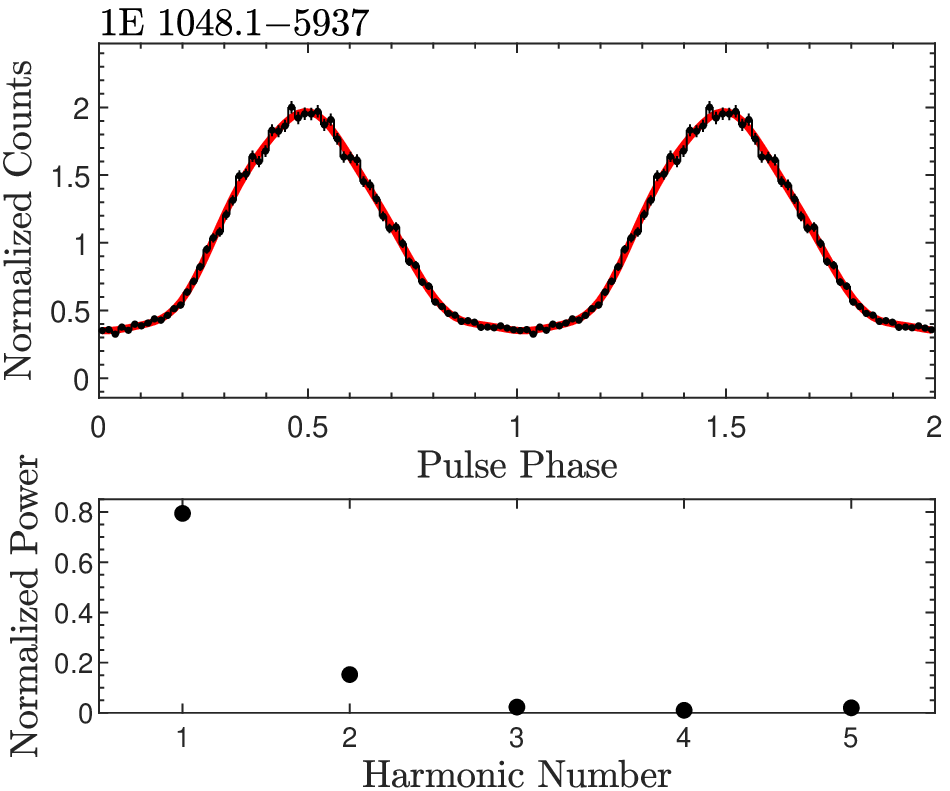}
\includegraphics[width=1.05\textwidth]{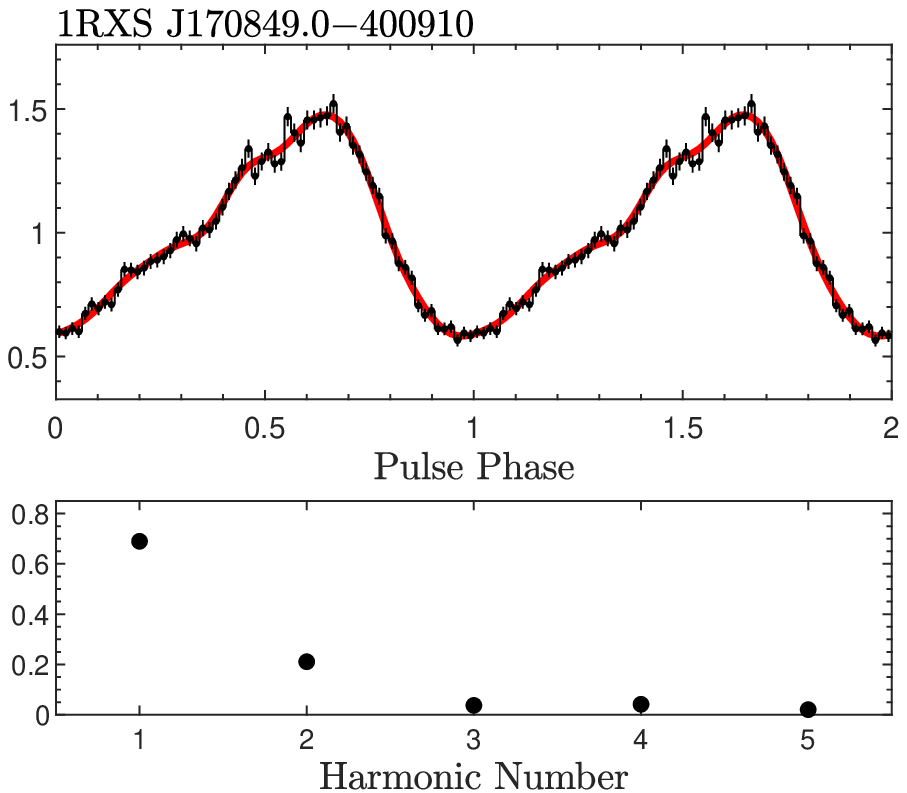}
\includegraphics[width=1.05\textwidth]{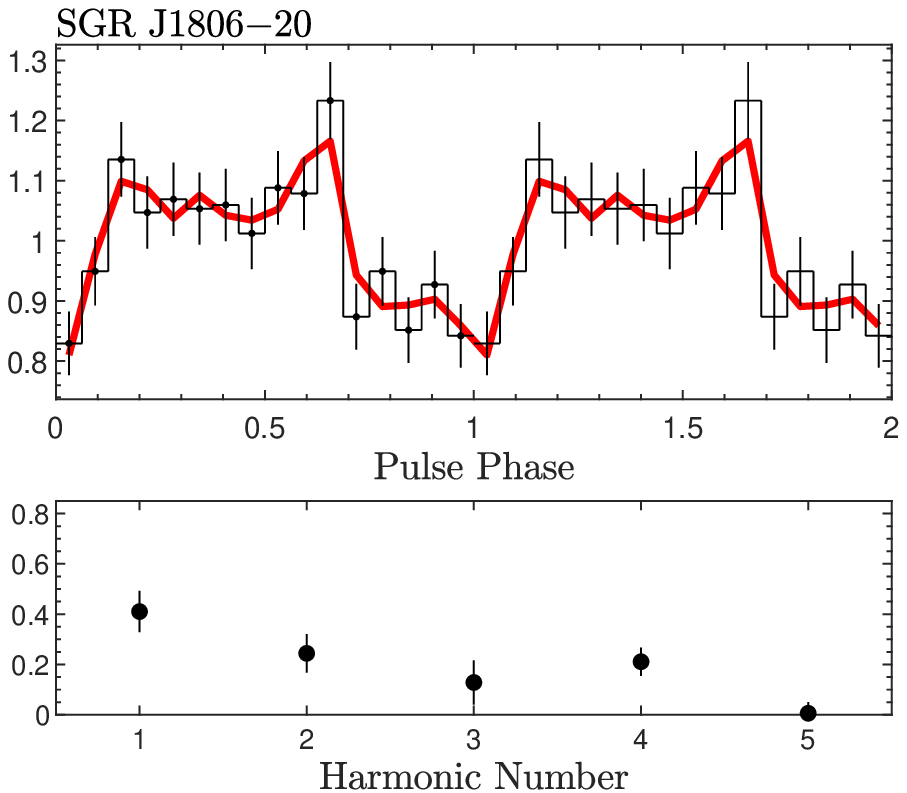}
\includegraphics[width=1.05\textwidth]{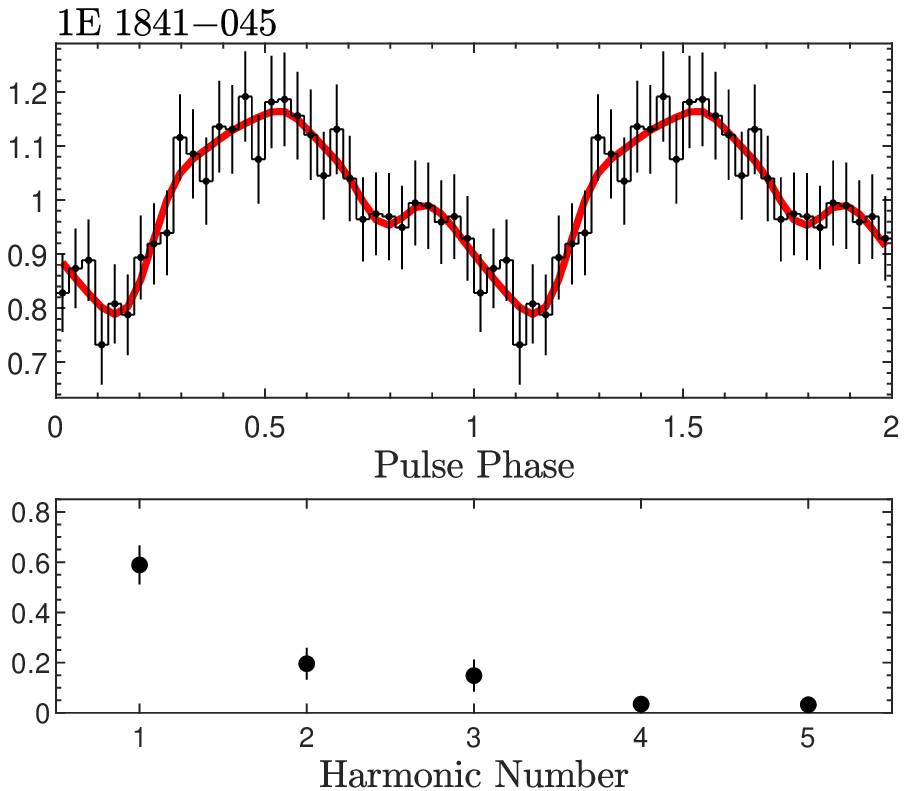}
\end{minipage}
\begin{minipage}{0.32\linewidth}
\includegraphics[width=1.05\textwidth]{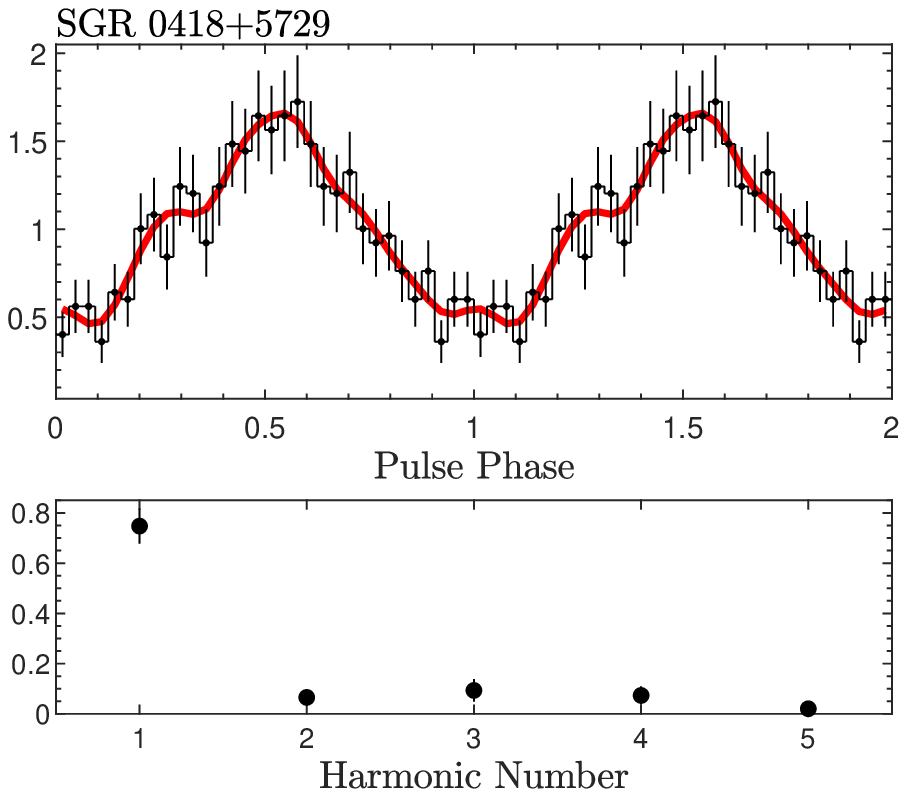}
\includegraphics[width=1.05\textwidth]{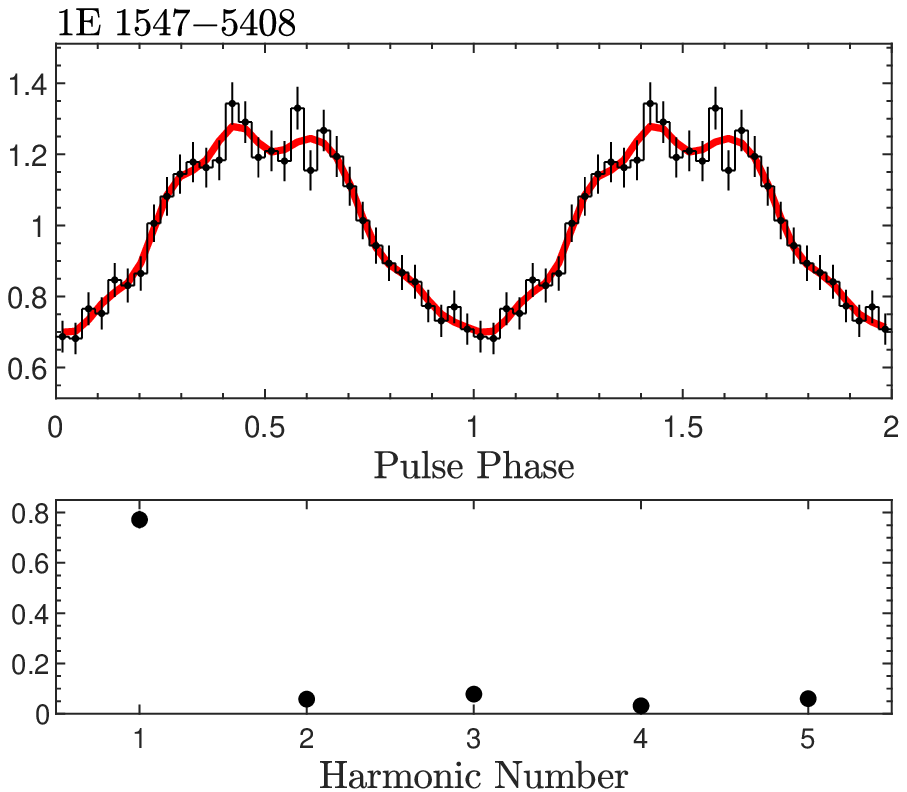}
\includegraphics[width=1.05\textwidth]{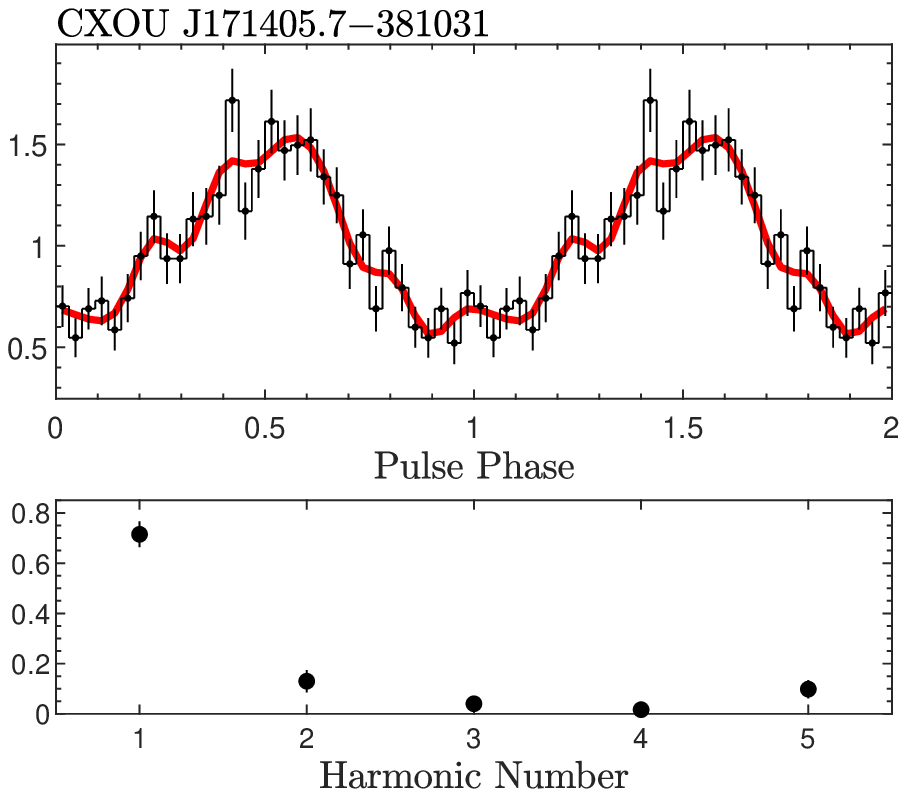}
\includegraphics[width=1.05\textwidth]{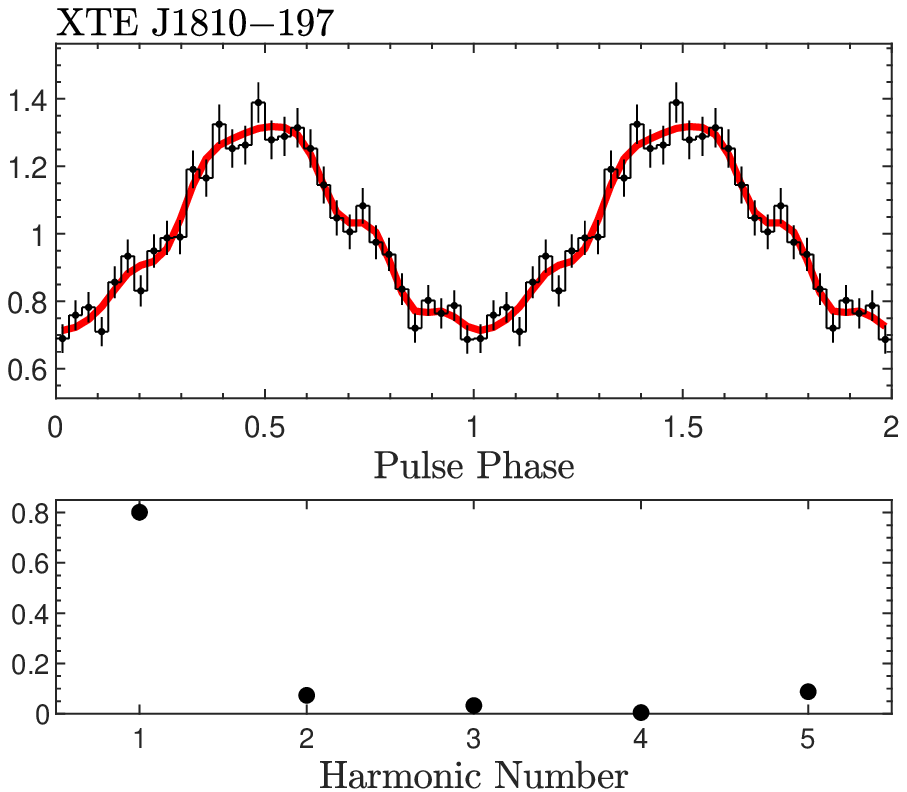}
\includegraphics[width=1.05\textwidth]{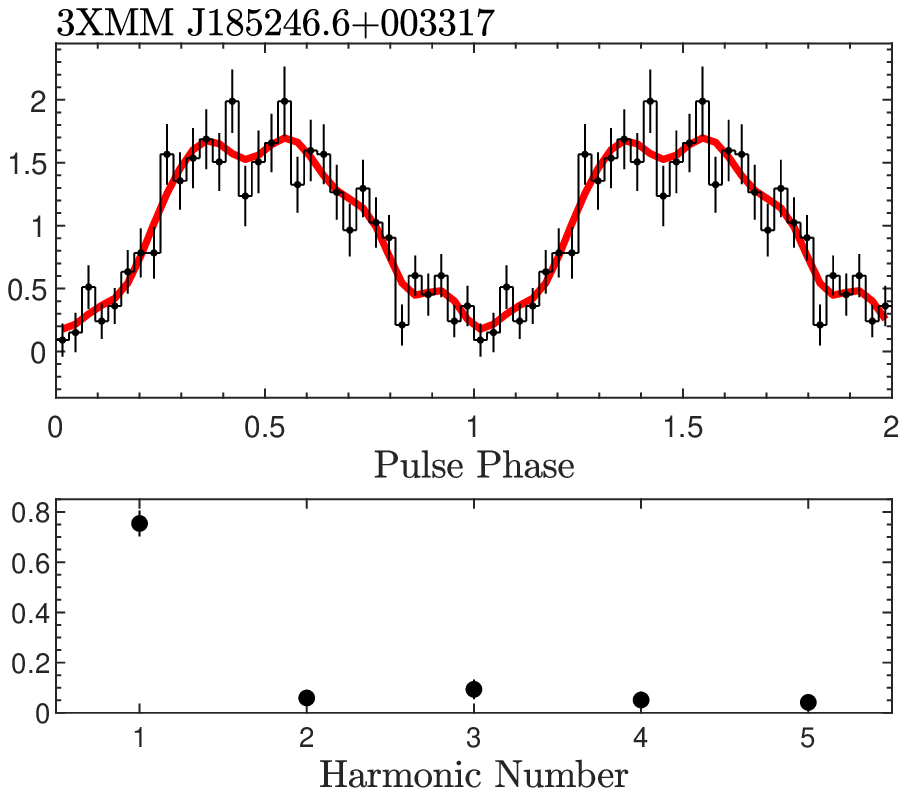}
\end{minipage}
\end{center}
\end{figure*}

\begin{figure*}
\begin{center}
\begin{minipage}{0.32\linewidth}
\includegraphics[width=1.05\textwidth]{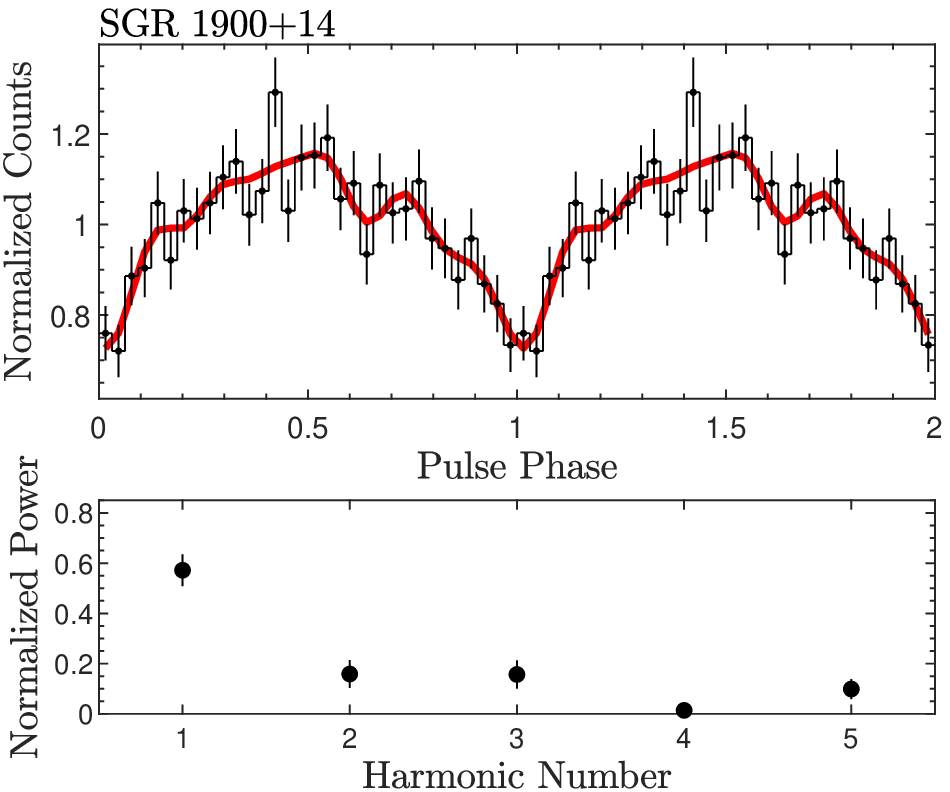}
\end{minipage}
\begin{minipage}{0.32\linewidth}
\includegraphics[width=1.05\textwidth]{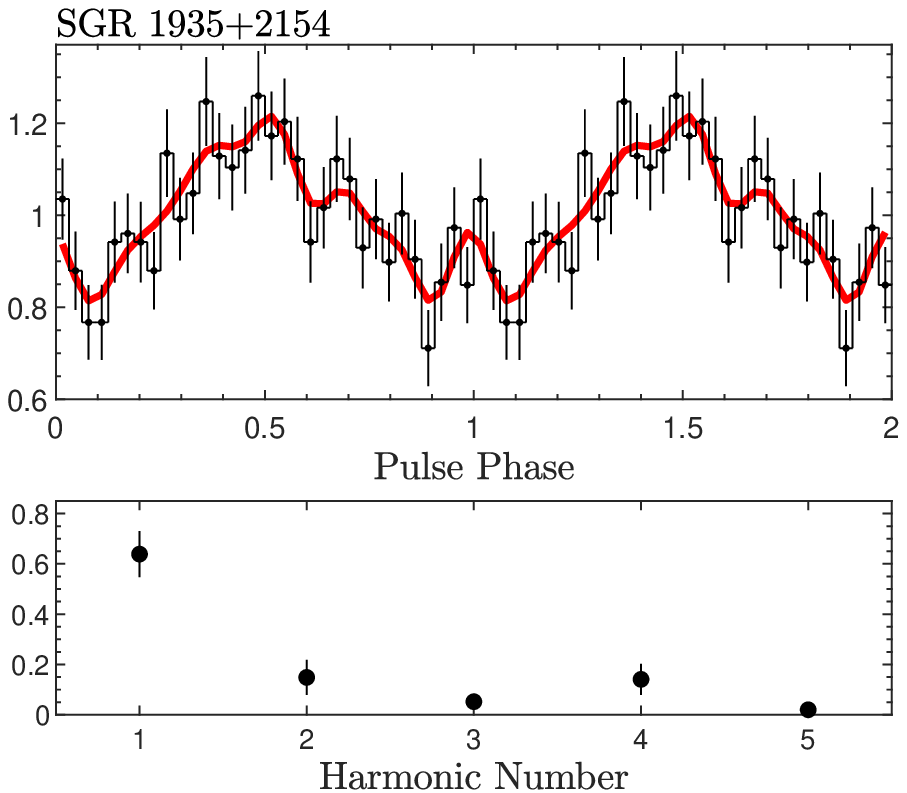}
\end{minipage}
\begin{minipage}{0.32\linewidth}
\includegraphics[width=1.05\textwidth]{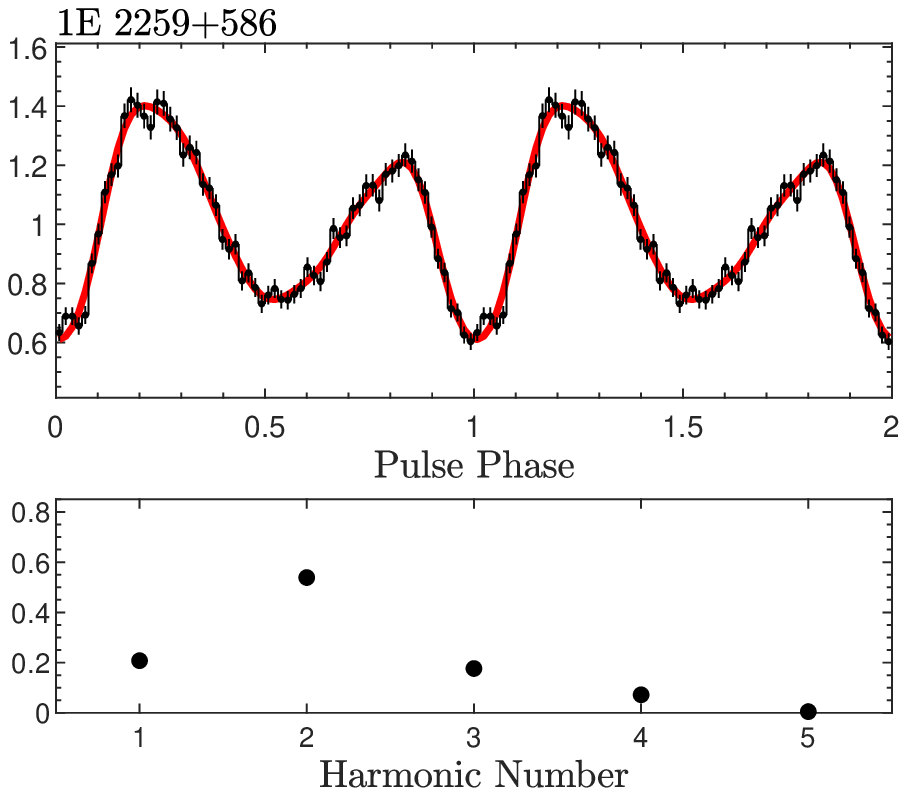}
\end{minipage}
\end{center}
\caption{Soft X-ray profiles (0.5--2\,keV) of sampled magnetars and the corresponding strength of Fourier components. The histogram and red curve in the upper panel of each source denote the observed pulse profile and the reconstructed one from the first five Fourier harmonics. The pulse profile of each source is normalized to its mean photon counts. The bin sizes shown here vary between sources and are chosen purely for illustrative purposes. The normalized Fourier power of the first five harmonics is plotted in the lower panel for each source. \label{profile}}
\end{figure*}

\begin{table*}
	\centering
	\caption{Physical properties and derived parameters of the sampled magnetars.}
	\label{summary_table}
	\begin{threeparttable}
	\begin{tabular}{lccccccccc} 
		\hline
		Name & $P^a$ & $-\log{\dot{P}}^a$ & $B$-Field$^a$ & $\tau_c^a$ & SNR Age & $\log{L}^b$ & $A_1$ &$A_2$ &PF$^c$ \\ 
 & (s) &  & ($10^{14}$\,G) & (kyr) & (kyr) & (\lumcgs) &  &  & \\
		\hline
CXOU~J0100 & 8.02 & 10.72 & 3.9 & 6.8 & -- & $35.4^{+0.4}_{-0.3}$ & $0.40\pm0.04$ & $0.30\pm0.04$ & $0.20\pm0.01$ \\
4U~0142 & 8.69 & 11.69 & 1.3 & 68 & -- & $35.5^{+0.2}_{-0.3}$ & $0.45\pm0.02$ & $0.42\pm0.02$ & $0.047\pm0.001$ \\
SGR~0418 & 9.08 & 14.40 & 0.061 & 36000 & -- & $31.0_{-0.8}^{+0.6}$ & $0.64\pm0.09$ & $0.09\pm0.05$ & $0.37\pm0.03$ \\
SGR~0501 & 5.76 & 11.22 & 1.9 & 16 & 4--7 & $33.5_{-1.1}^{+0.7}$ & $0.38\pm0.03$ & $0.28\pm0.03$ & $0.28\pm0.01$ \\
1E~1048 & 6.46 & 10.65 & 3.9 & 4.5 & -- & $34.8\pm0.2$ & $0.79\pm0.01$ & $0.152\pm0.006$ & $0.584\pm0.004$ \\
1E~1547 & 2.07 & 10.32 & 3.2 & 0.69 & -- & $33.1_{-0.8}^{+0.4}$ & $0.73\pm0.06$ & $0.07\pm0.03$ & $0.205\pm0.009$ \\
CXOU~J1647 & 10.61 & 12.01 & 1.0 & 173 & -- & $33.8_{-0.7}^{+0.4}$ & $0.63\pm0.08$ & $0.14\pm0.05$ & $0.47\pm0.04$ \\
1RXS~J1708 & 11.01 & 10.71 & 4.7 & 9.0 & -- & $34.7\pm0.2$ & $0.69\pm0.02$ & $0.21\pm0.01$ & $0.294\pm0.004$ \\
CXOU~J1714 & 3.83 & 10.19 & 5.0 & 0.95 & $0.65^{+2.50}_{-0.30}$ & $34.9^{+0.7}_{-1.1}$ & $0.63\pm0.07$ & $0.13\pm0.04$ & $0.31\pm0.02$ \\
SGR~J1745 & 3.76 & 10.86 & 2.3 & 4.3 & -- & $34.2_{-0.3}^{+0.2}$ & $0.62\pm0.07$ & $0.07\pm0.04$ & $0.26\pm0.03$ \\
SGR~1806 & 7.55 & 9.31 & 20 & 0.24 & -- & $35.3_{-0.3}^{+0.2}$ & $0.41\pm0.08$ & $0.24\pm0.07$ & $0.10\pm0.02$ \\
XTE~J1810 & 5.54 & 11.11 & 2.1 & 11 & -- & $34.6_{-0.3}^{+0.2}$ & $0.74\pm0.06$ & $0.08\pm0.03$ & $0.212\pm0.009$ \\
Swift~J1822 & 8.44 & 13.68 & 0.14 & 6300 & -- & $33.0_{-0.9}^{+0.6}$ & $0.69\pm0.06$ & $0.14\pm0.03$ & $0.33\pm0.02$ \\
1E~1841 & 11.79 & 10.39 & 7.0 & 4.6 & 0.5--1 & $35.3_{-0.5}^{+0.4}$ & $0.50\pm0.09$ & $0.19\pm0.07$ & $0.11\pm0.02$ \\ 
3XMM~J1852& 11.56 & $>$12.85 & $<$0.41 & $>$1300 & -- & $33.7_{-0.6}^{+0.4}$ & $0.67\pm0.07$ & $0.09\pm0.04$ & $0.52\pm0.04$ \\
SGR~1900 & 5.20 & 10.04 & 7.0 & 0.9 & -- & $35.0_{-0.3}^{+0.2}$ & $0.50\pm0.08$ & $0.15\pm0.06$ & $0.11\pm0.01$ \\
SGR~1935 & 3.25 & 10.87 & 2.2 & 3.6 & -- & $34.2_{-0.6}^{+0.4}$ & $0.5\pm0.01$ & $0.13\pm0.06$ & $0.10\pm0.02$ \\
1E~2259 & 6.98 & 12.32 & 0.59 & 230 & $14\pm0.02$ & $34.8\pm0.2$ & $0.21\pm0.01$ & $0.53\pm0.02$ & $0.233\pm0.005$\\
		\hline
	\end{tabular}
	\begin{tablenotes}
            \item[a] $P$, $\dot{P}$, $B$ field, $\tau_c$, and $\tau_{SNR}$ are adopted from McGill Online Magnetar Catalog \citep{OlausenK2014}. The timing solution of CXOU~J1647 is from the measurement by \citep{RodriguezIE2014}. We further updated the values for CXOU~J1714 according to our new measurements (see Appendix \ref{new_pdot}).
            \item[b] Bolometric thermal luminosity estimated from the best-fit spectral parameters in \citet{MongN2017} except for SGR 0418 \citep{ReaIP2013}, SGR~1745 \citep{ZelatiRT2017}, 3XMM~J1852 \citep{ZhouCL2014}, and SGR~1935 (this work). The distance uncertainties were taken into account. We assumed a relative error of 50\,\% if the distance uncertainty is not reported.
            \item[c] The uncertainties of $A_1$, $A_2$, and PF denote 1-$\sigma$ confidence interval.
        \end{tablenotes}
        \end{threeparttable}
\end{table*}

\section{Analysis Method}\label{method}
Given that we are focusing on the quiescent soft X-ray pulse profile, we chose to analyse the events in the energy range of 0.5--2\,keV. SGR 1745$-$2900 and SGR 1806$-$20 are exceptional due to insufficient photon count (see Table \ref{summary_dataset}). Although a precise spin
period is needed to investigate the detailed pulse-profile structures, some new observations are not covered by published ephemerides \citep[see e.g.,][]{DibK2014}. Moreover, many magnetars have no long-term ephemerides.We therefore searched the periods for individual data set using the $H$-test algorithm \citep{deJagerRS1989} and obtained their soft X-ray profiles, in which the emission is dominated by the surface thermal emission. For magnetars with multiple observations, we searched their periods individually.We verified that the profile did not change significantly between observations. We then assigned the pulse phase to each photon with respect to a fiducial point, for example the valley of the pulse. We combined all the photon events to create a stacked pulse profile. The pulse profile could have minor variability between observations as a result of noise. However, we checked that there were no significant variabilities (e.g. from single-peaked to double-peaked). We divided the pulse profile into 64 phase bins. We also performed our analysis using 16 and 32 bins, but found that the results were consistent across each choice of binning. We calculated the uncertainty of each phase bin in the folded light curve by assuming a Poisson distribution for the photons. The background folded light curve was created according to the same timing solution and then was subtracted from the source profile. The background-subtracted pulse profiles of sampled magnetars are plotted in Fig.~\ref{profile}.

We only collected $\sim$30 X-ray photons in 0.5--2\,keV from SGR J1745$-$2900 owing to the heavy absorption \citep{ZelatiRP2015}. We then calculated the $H$ value for different energy bands. For each calculation, we set the low-energy boundary to 0.5\,keV and let the high energy boundary $E_h$ vary from 2 to 7\,keV. We found that the $H$ value increases monotonically starting from $E_h=3$\,keV and reaches a plateau at $E_h\gtrsim5.5$\,keV. The pulse profile in 0.5--4\,keV results in $H=63$, corresponding to a detection significance of 6.5$\sigma$ \citep{deJager2010}. We therefore calculated the parameters of the pulse profile at $\lesssim4$\,keV. Similarly, the X-ray photons collected from SGR 1806$-$20 below 2\,keV are insufficient for a timing analysis. In addition, the pulse profile above 4\,keV could be very different from that below 4\, keV \citep{YounesKK2015}. We therefore extended the high-energy boundary to 4\,keV to increase the X-ray photon numbers.

We applied the Fourier transform on the pulse profiles and measured the amplitude of each harmonic to quantify the profile shape. We only considered harmonic numbers $k<6$ because the higher-order terms have negligible power. We then calculated the relative strength of each harmonic by
\begin{equation}
A_k=\frac{\sqrt{a_k^2+b_k^2}}{\sum_{j=1}^5\left( \sqrt{a_j^2+b_j^2} \right)},
\end{equation}
for $k=$1--5, where
\[ a_k=\frac{1}{N}\sum_{i=1}^N x_i\cos\left( 2\pi k \phi_i \right)
\] 
and
\[ b_k=\frac{1}{N}\sum_{i=1}^N x_i\sin\left( 2\pi k \phi_i \right)
\]
are Fourier amplitudes of each component, $x_i$ is the number of photons in the $i$th bin, $\phi$ is the phase of the $i$th bin, and $N$ is the total number of bins. We derived the uncertainties of the amplitudes with Monte Carlo simulations. We created $10^4$ pulse profiles based on the observed data points plus Gaussian-distributed random numbers with the standard deviation equal to the uncertainties of each bin. We calculated the amplitude of the Fourier components of all the simulated pulse profiles. They are well represented as a normal distribution, and hence we took the standard deviation as the 1$\sigma$ uncertainty. 

Except for the modulation shape, we further calculated the pulsed fraction (PF) of each magnetar. We employed the definition of the root mean square (rms) pulse amplitude based on the Fourier decomposition \citep{DibKG2009, AnAH2015}. PF is defined as 
\begin{equation}
\textrm{PF}=\frac{1}{a_0} \sqrt{2\sum_{k=1}^5 \left[ \left(a_k^2+b_k^2\right)-\left(\sigma_{a_k^2}+\sigma_{b_k^2}\right) \right]},
\end{equation}
where 
\[
\sigma_{a_k}^2=\frac{1}{N^2}\sum_{i=1}^N \sigma_i^2 \cos^2 \left( 2\pi k \phi_i \right)
\]
and
\[
\sigma_{b_k}^2=\frac{1}{N^2}\sum_{i=1}^N \sigma_i^2 \sin^2 \left( 2\pi k \phi_i \right)
\]
are the Fourier power generated by the noise, and $\sigma_i$ is the uncertainty of $x_i$. Compared with the conventional PF definition based on the area under the profile, this rms definition is less biased for data sets with large uncertainties and more suitable for complex profiles \citep[see][for more discussion about the characteristics of different PF definitions]{AnAH2015}. 

\section{Results}\label{result}
\subsection{Distribution of Profile Parameters}
We applied the above analysis techniques to the selected data sets and summarize the PF values and the strengths of the first two Fourier components of sampled magnetar profiles in Table \ref{summary_table}. None of the magnetars in our sample shows significant triple-peaked or more complex profiles in quiescence. Therefore, the strengths of $A_3$--$A_5$ are generally negligible, and we do not list them in Table \ref{summary_table}. We also give the spin parameters ($P$ and $\dot{P}$), spin-down-inferred parameters ($B$ field and characteristic age $\tau_c$), and the age measurements from the supernova remnants (SNRs) if available. We also list their thermal luminosities in the table (see Appendix \ref{spectral_analysis}). 
\begin{figure}
\begin{center}
\includegraphics[width=0.49\textwidth]{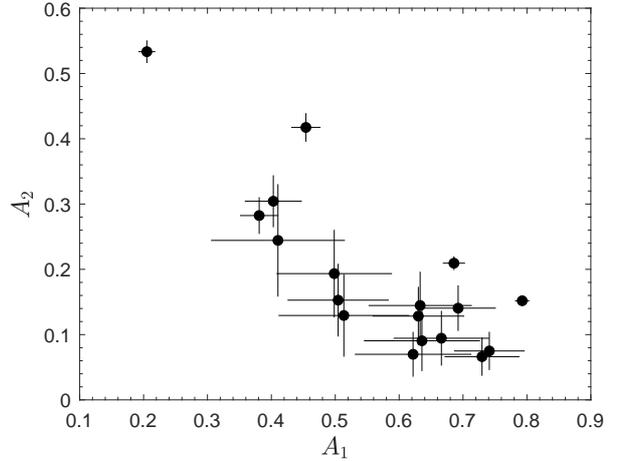}
\end{center}
\caption{Correlation between $A_1$ and $A_2$. \label{a1_a2}}
\end{figure}

\begin{figure*}
\begin{center}
\begin{minipage}{0.49\linewidth}
\includegraphics[width=0.95\textwidth]{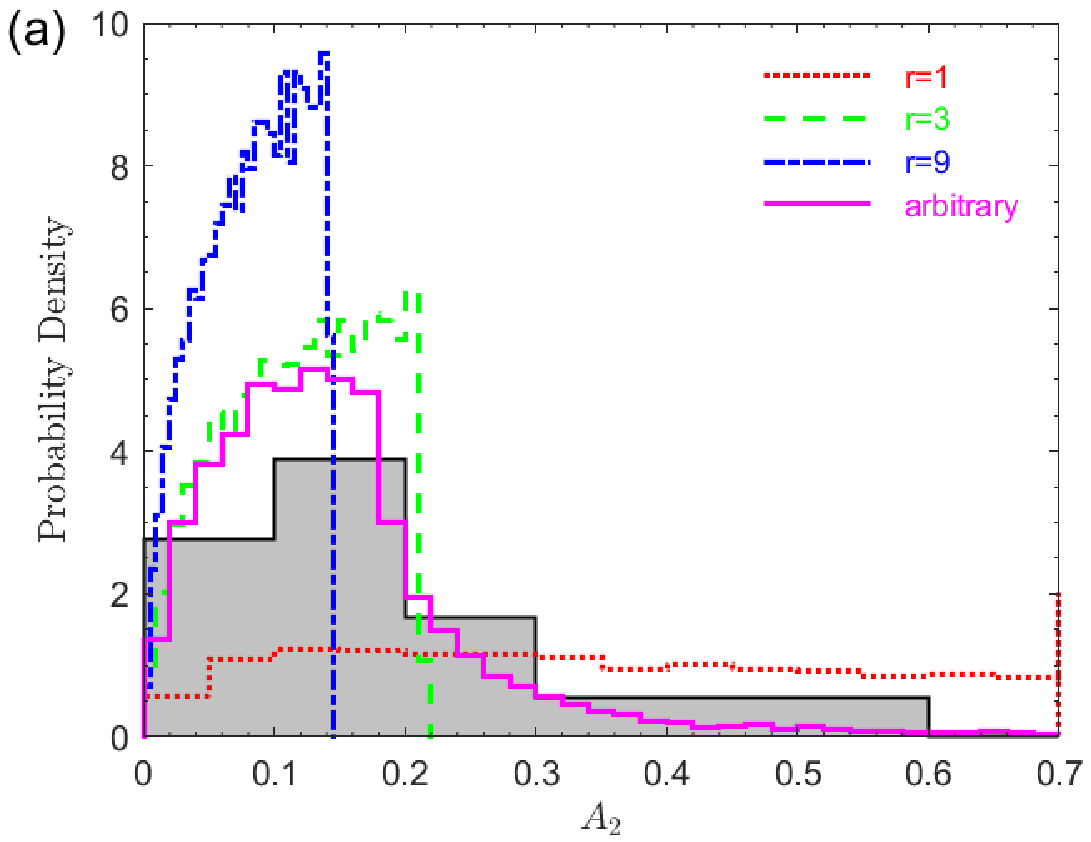}
\end{minipage}
\begin{minipage}{0.49\linewidth}
\includegraphics[width=0.95\textwidth]{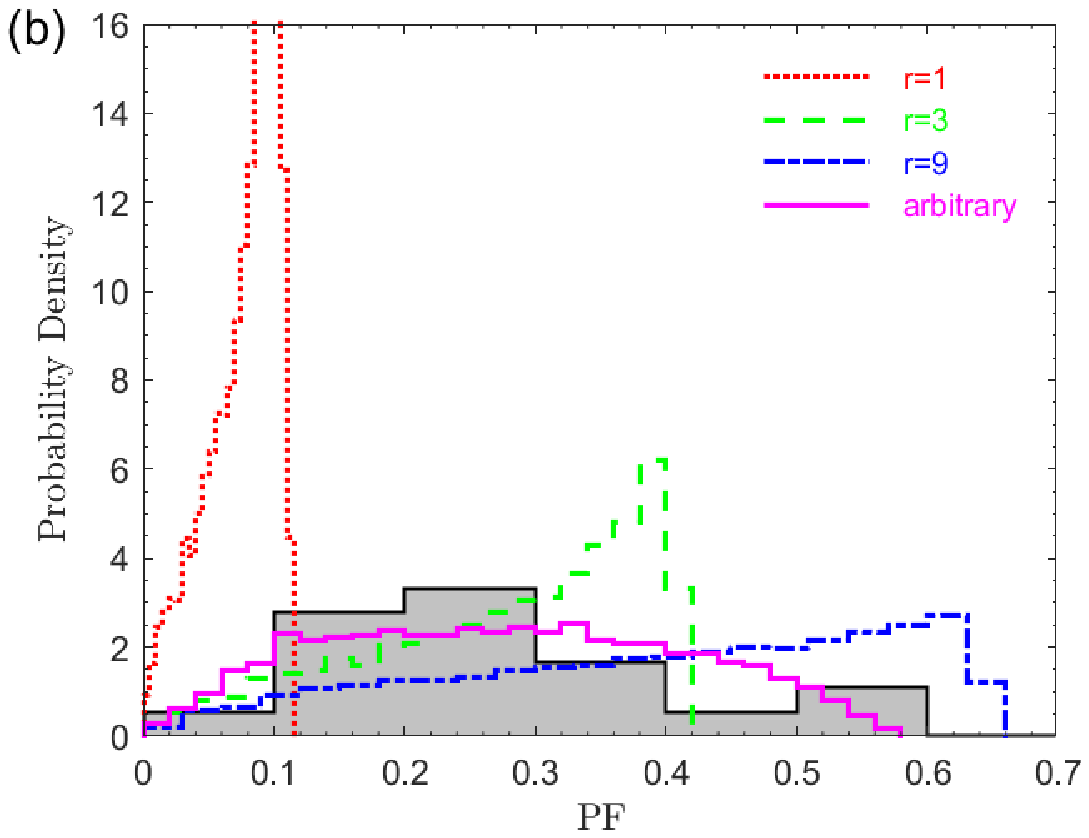}
\end{minipage}
\end{center}
\caption{Distributions of (a) $A_2$ and (b) PF for observed and simulated pulse profiles. The grey histograms are the observed distributions. We overplot the histograms of simulated pulsed profiles for two hotspots with intensity ratios of $r=1$ (red dotted line), $r=3$ (green dashed line), $r=9$ (blue dashed-dotted line), and arbitrary intensity ratios between 1 and 6 (solid magenta line).  \label{a2p_pf_distribution}}
\end{figure*}

\begin{figure}
\includegraphics[width=0.49\textwidth]{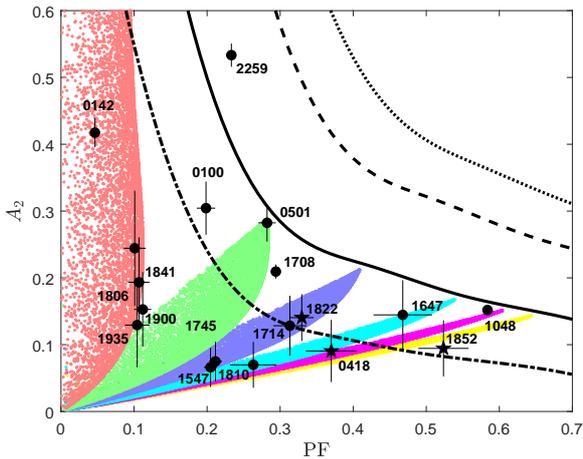}
\caption{$A_2$ versus PF for observed and simulated profiles. The dots and stars are observed values (see Table \ref{summary_table}), with the latter indicating low-$B$-field magnetars.  Error bars denote 1$\sigma$ uncertainties. The red, green, blue, cyan, magenta and yellow points are the simulated data points for intensity ratios between two hotspots of 1 (symmetric), 2, 3, 5, 7 and 9, respectively. The beaming is calculated using the Hopf function. The thick solid line represents the maximum available $A_2$ and PF for the given Hopf beaming. The dotted, dashed, and dash–dotted lines represent strong beaming with $\cos^3\theta'$, the beaming function adopted from \citet{AdelsbergL2006}, and the isotropic emission, respectively  (see Appendix \ref{modeling}). \label{a2p_pf_allcase}}
\end{figure}

Because $A_3$--$A_5$ are generally negligible, our magnetar sample shows an anticorrelation between $A_1$ and $A_2$ (see Fig.~\ref{a1_a2}). Pearson’s linear correlation coefficient is $-0.82$, with a null hypothesis probability of $3.2\times 10^{-5}$. Therefore, we focus on $A_2$ and PF in the following analysis and discussion. Four magnetars, CXOU~J0100, 4U~0142, SGR~0501, and 1E~2259, clearly show double-peaked profiles. They have large $A_2\gtrsim0.3$ and small $A_1\lesssim0.4$. The other 14 magnetar profiles are single-peaked and have weak $A_2\lesssim0.2$. The distribution of $A_2$ and PF is shown in Fig.~\ref{a2p_pf_distribution}. It is clear that $A_2$ exhibits a peak at $0.15$ and a skewed tail extending to $0.6$ (Fig.~\ref{a2p_pf_distribution}a). The contributions to the tail come from the four above-mentioned magnetars with double-peaked profiles. In contrast, PF shows a flatter distribution, with a peak located at $0.25$. Three magnetars, 1E~1048, 3XMM~J1852, and CXOU~J1647, have the highest $\textrm{PF}\gtrsim0.5$, and they also systematically show single-peaked profiles with low $A_2$. We then investigated the connection between $A_2$ and $PF$ (Fig.~\ref{a2p_pf_allcase}).  Because of the anticorrelation between $A_1$ and $A_2$ seen in Fig.~\ref{a1_a2}, PF versus $A_1$ would be essentially a simple inversion of the plot shown in Fig.~\ref{a2p_pf_allcase}. Obviously, magnetars are not uniformly distributed in the PF-$A_2$ plot. More than half of the magnetars have low PF$\lesssim0.3$ and low $A_2\lesssim0.3$. Others have high PF but low $A_2$, or vice versa. No magnetar shows both high $A_2$ and high PF.

\subsection{Correlation with Physical Parameters}

\begin{figure*}
\begin{center}
\begin{minipage}{0.49\linewidth}
\includegraphics[width=0.95\textwidth]{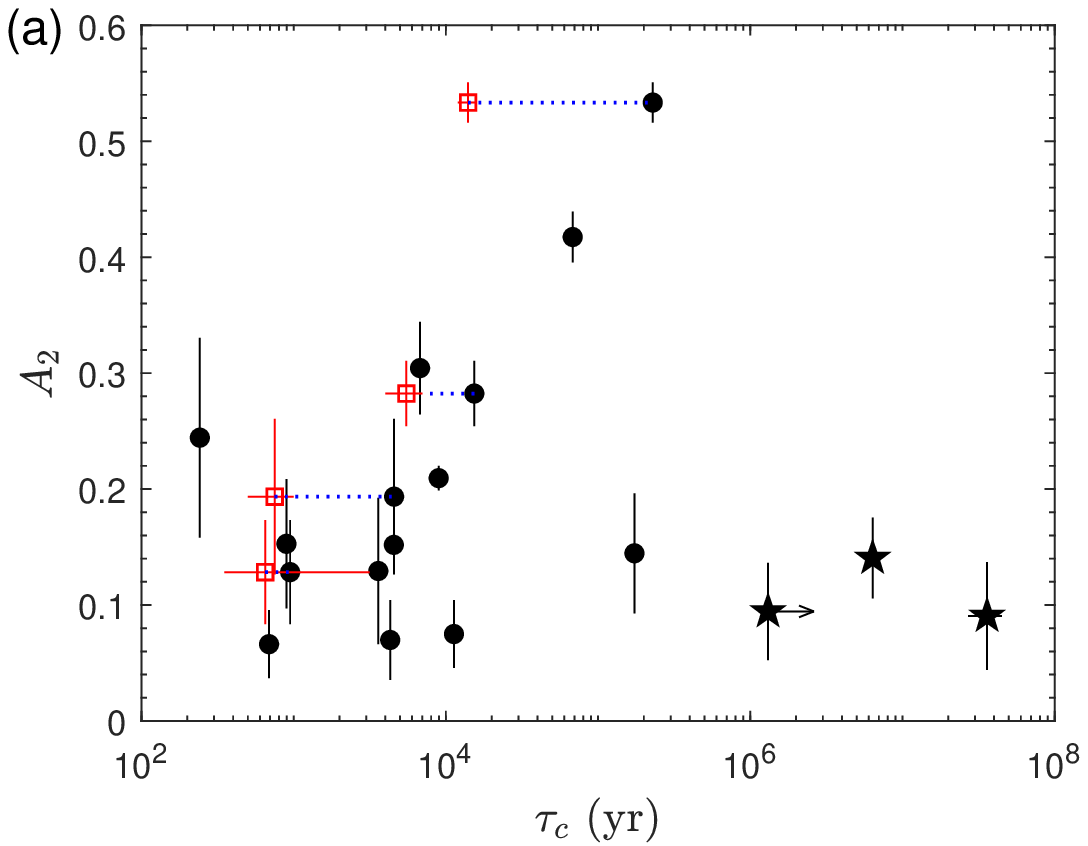}
\includegraphics[width=0.95\textwidth]{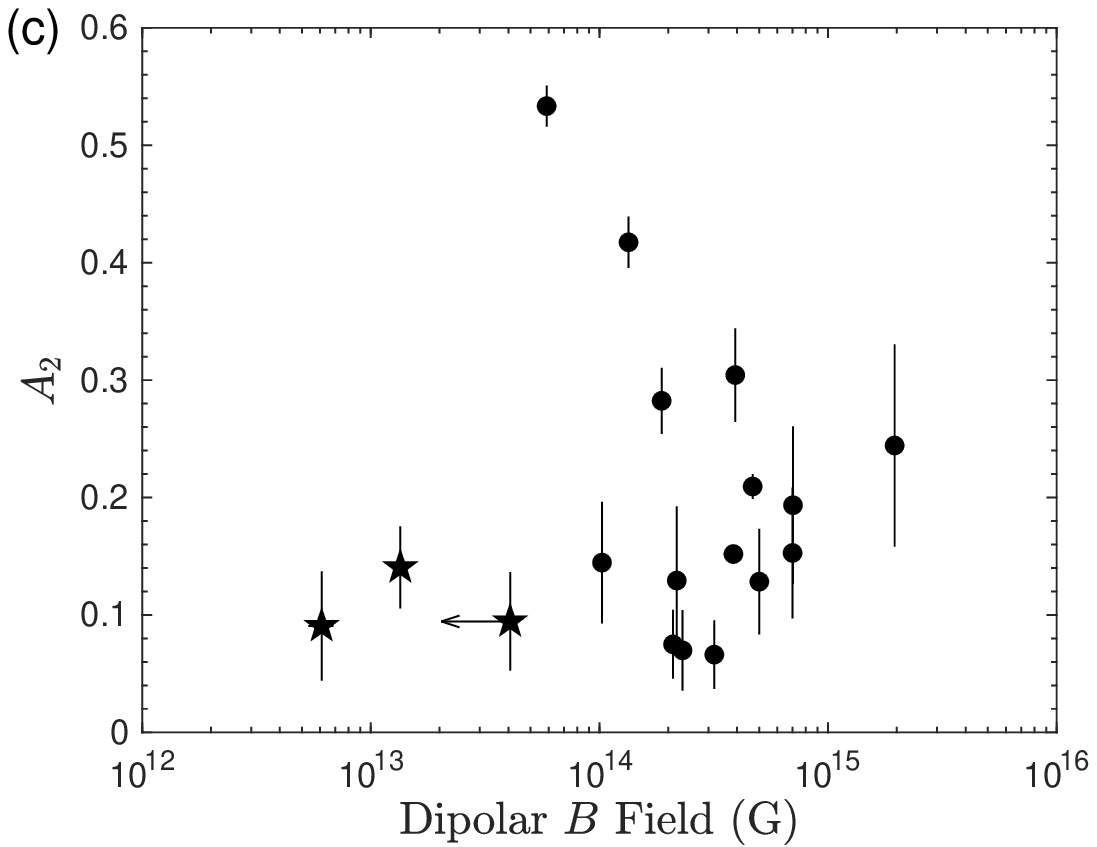}
\includegraphics[width=0.95\textwidth]{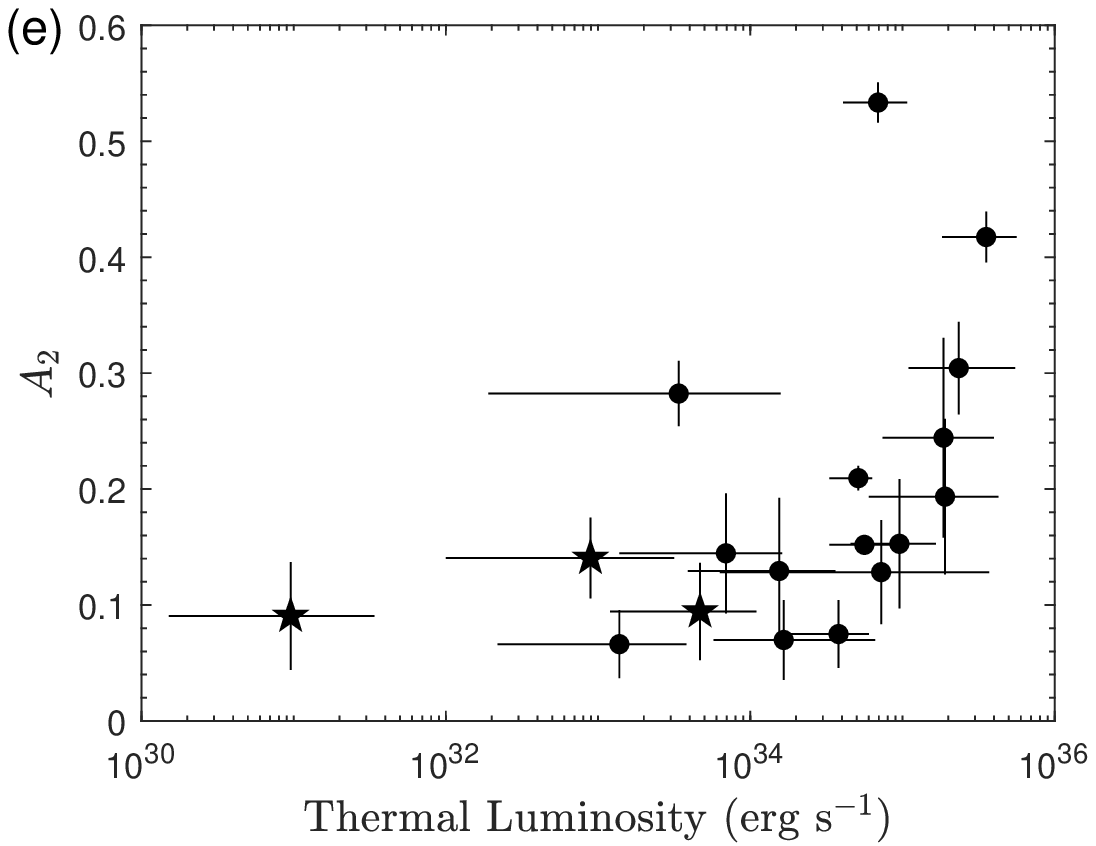}
\end{minipage}
\begin{minipage}{0.49\linewidth}
\includegraphics[width=0.95\textwidth]{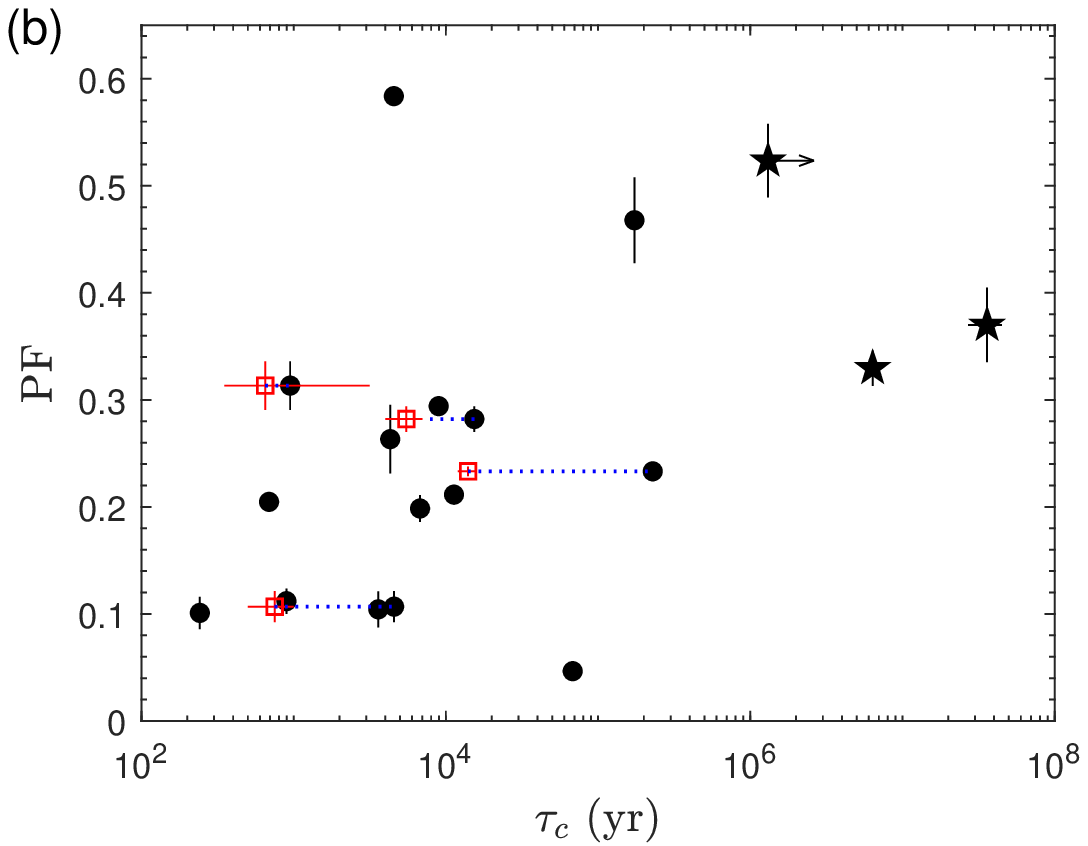}
\includegraphics[width=0.95\textwidth]{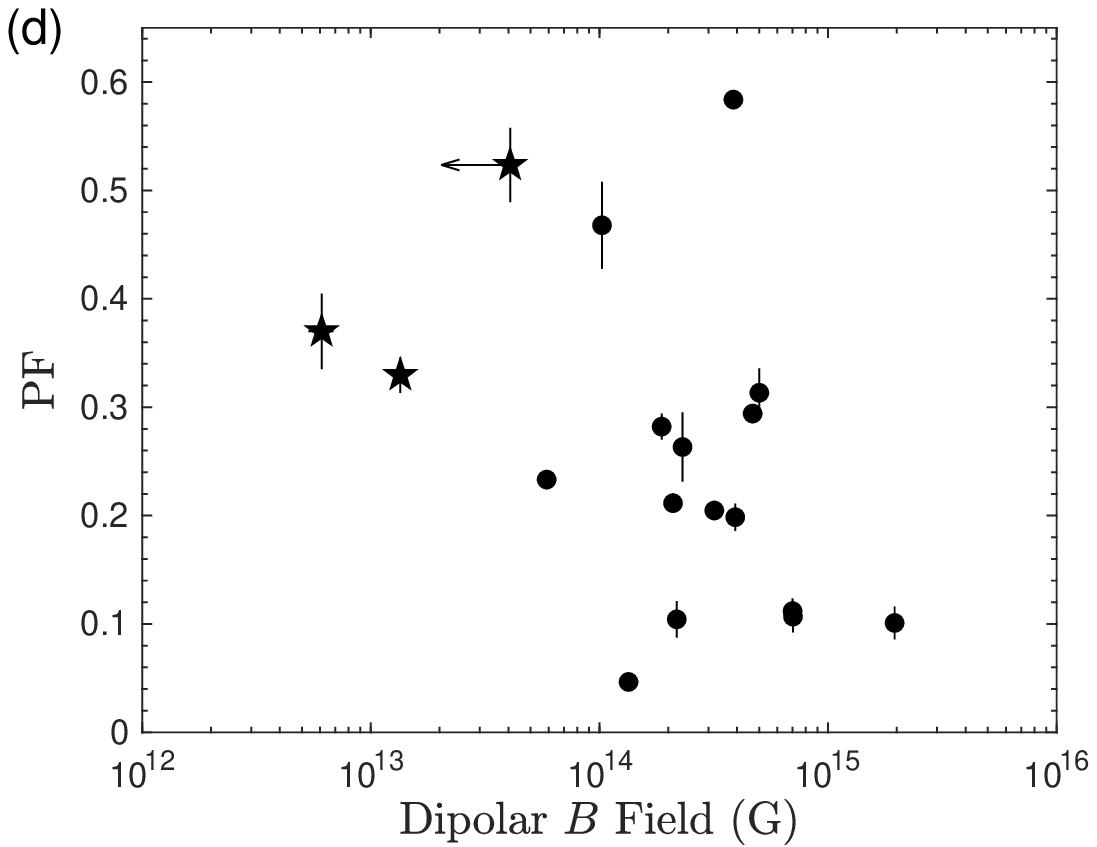}
\includegraphics[width=0.95\textwidth]{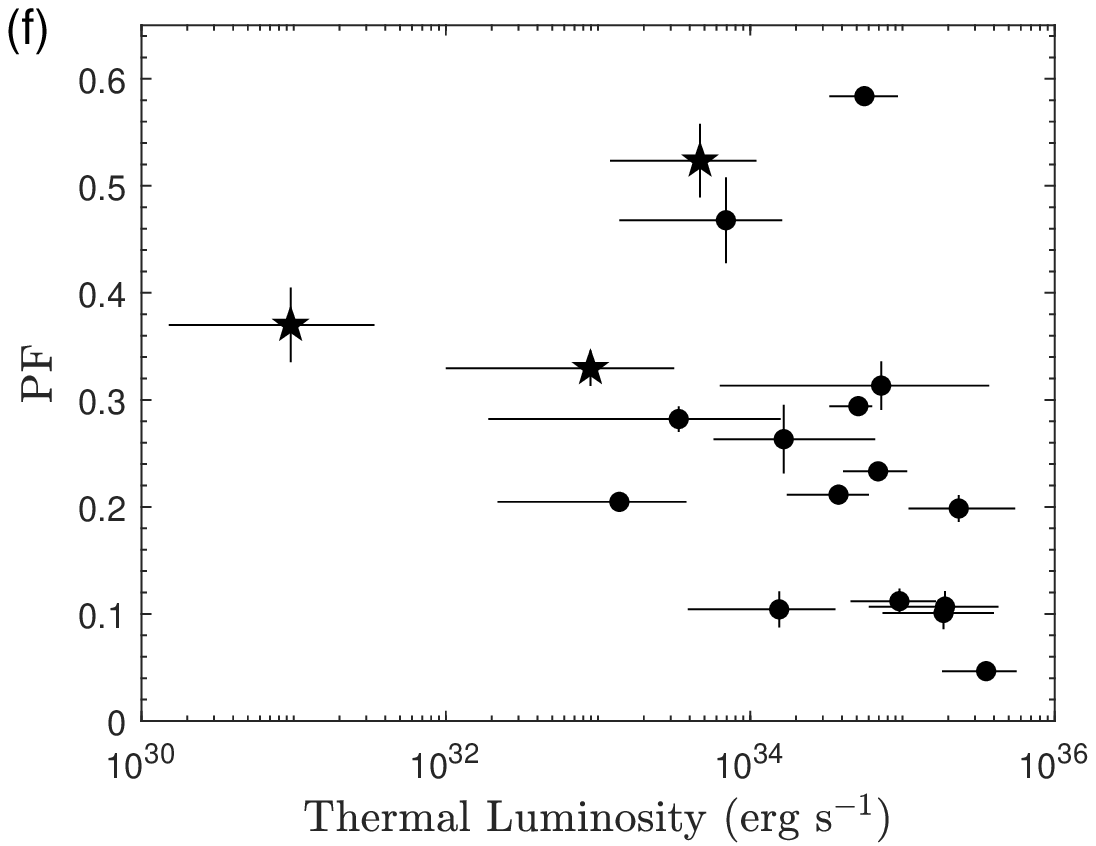}
\end{minipage}
\end{center}
\caption{$A_2$ and PF versus $\tau_c$, the $B$ field, and the thermal luminosity. The star symbols refer to the low-$B$-field magnetars. The error bars denote 1$\sigma$ uncertainties for $A_2$ and PF. The uncertainties of the thermal luminosity are from the 90 per cent confidence interval through spectral fitting in the literature and the distance measurements. They are dominated by the assumed 50 per cent distance uncertainty (see text). Four magnetars with well-measured SNR ages (CXOU~J171405.7$-$381031, 1E~1841$-$045, SGR~0501+4516, and 1E~2259+586) are further denoted by red squares and connected to their $\tau_c$ with blue dotted lines. \label{age_parameter}}
\end{figure*}

We further investigate the correlation between pulse-profile parameters and physical parameters. Figs \ref{age_parameter}(a) and (b) show the age dependence of $A_2$ and PF. Four magnetars, CXOU~J1714, 1E~1841, SGR~0501, and 1E~2259, have well measured SNR ages that are much younger than their $\tau_c$ \citep{LeahyT2007, NakamuraBI2009, TianL2008, SasakiPG2013}. We plotted them on the same figure for comparison. $A_2$ has no clear correlation. On the other hand, PF has an intriguing age dependence. However, the
correlation analysis does not yield a significant linear correlation, with a null hypothesis probability of 0.4. Considering that the correlation could be non-linear, we calculated Spearman’s rank correlation and found a coefficient of 0.5 with a null hypothesis probability of 0.04, only slightly higher than 2$\sigma$ significance. We further plotted the $B$-field dependence of $A_2$ and PF in Figs \ref{age_parameter}(c) and (d) to check if there is any correlation, although the spin-down inferred $B$ field contains only the dipolar term. The distributions are similar to a horizontal flip of those with respect to $\tau_c$ because the spin period of magnetars is distributed in a narrow range \citep[see e.g.,][]{Ho2013}. Both the Pearson and the Spearman correlation analysis yield a weak anti-correlation between PF and $B$ field with a coefficient of $\sim-0.5$ and a null hypothesis probability of $0.05$ (Pearson) and $0.01$ (Spearman). 

Magnetars are believed to have complex $B$-field structures that cannot be inferred solely from spin-down. The thermal luminosity could provide hints about the hidden fields according to the magneto-thermal evolution model. Therefore, we examined the thermal luminosity dependence of $A_2$ and PF, as shown in Figs \ref{age_parameter}(e) and (f). We found a weak correlation between $A_2$ and the thermal luminosity, with a null hypothesis probability of $0.01$ (Pearson) and $0.008$ (Spearman). PF weakly anticorrelates with the thermal luminosity, and both correlation analysis methods suggest a correlation coefficient of $\sim -0.6$ and a null hypothesis probability of $0.01$. This is expected because the dipolar $B$ field of magnetars positively correlates with the X-ray luminosity \citep{AnKT2012, MongN2017}.

\begin{figure}
\begin{center}
\includegraphics[width=0.49\textwidth]{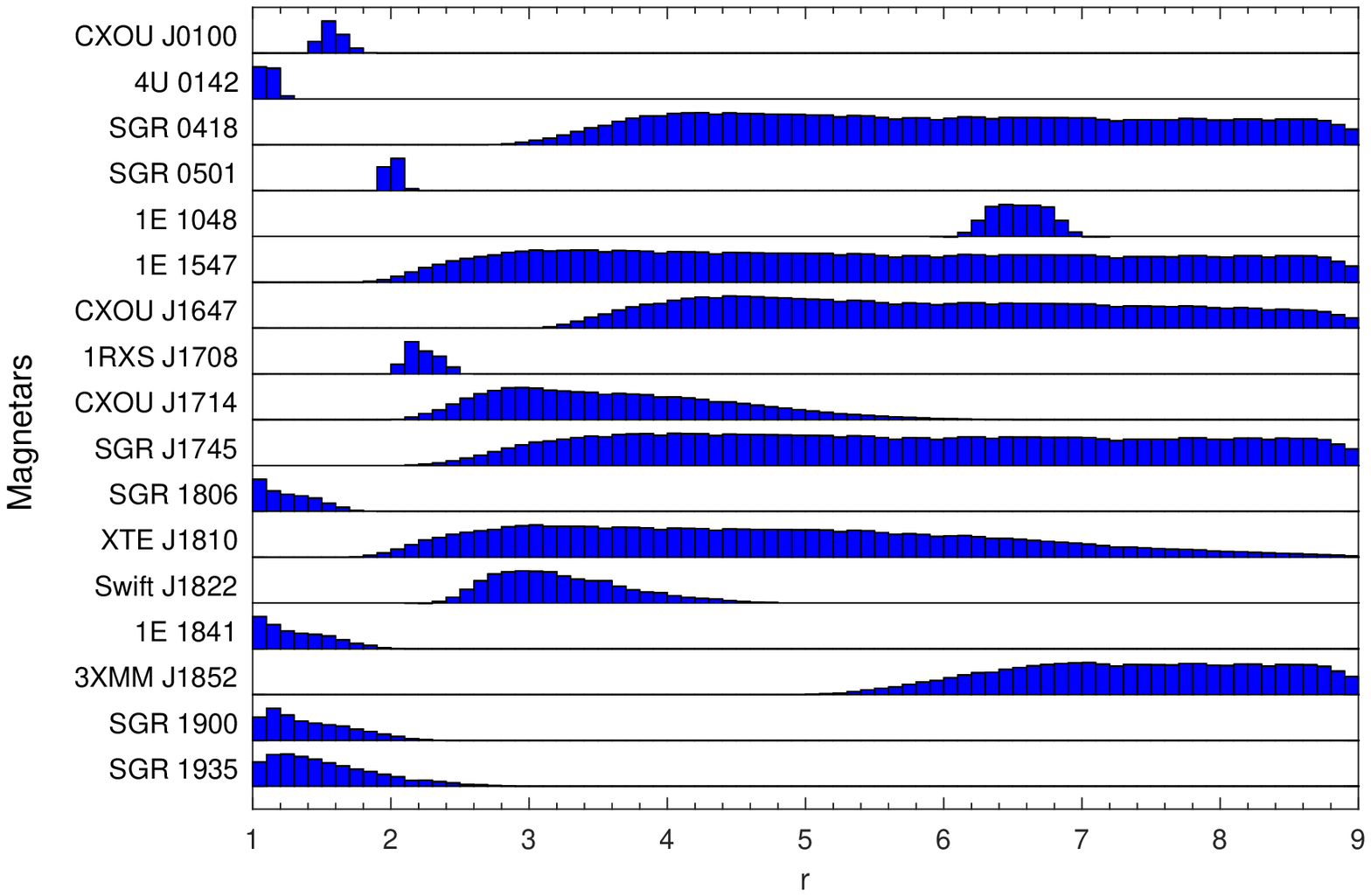}
\end{center}
\caption{Distribution of the intensity ratio between two hotspots for magnetars. The probability densities are scaled to have identical peak heights for display purposes.  \label{magnetar_ratio}}
\end{figure}

\section{Simulation of Pulse Profiles}\label{simulation}
In the soft X-ray band, the emission is dominated by the thermal radiation from the surface. The pulse profiles therefore reflect the anisotropy of the surface temperature distribution. Theoretical works suggest that the temperature profile could show localized hotspots or an asymmetric pattern \citep{ViganoRP2013, GourgouliatosWH2016}.  In practice, the surface temperature profile and the emission geometry of a few magnetars were constrained by using models consisting of two antipodal hotspots with different intensities, for example for SGR 0418+5729 \citep{GuillotPR2015}, XTE J1810$-$197 \citep{PernaG2008, BernardiniPG2011}, and PSR J1119$-$6127 \citep{NgKH2012}. We applied the same method as described in \citet{DedeoPN2001} to our sample of magnetars. We also note that the viewing geometry plays a critical role in the observed profiles. For two hotspots located at the magnetic poles, the observed pulse profile could be single-peaked, double-peaked, or even quadruple-peaked depending on the line of sight \citep[see e.g., ][]{Ho2007}. It is possible to test the surface temperature anisotropy of magnetars in a statistical way with simulations of different viewing geometries.

We first assumed a symmetric surface intensity distribution with respect to the magnetic equator, namely two antipodal hotspots with the same size and intensity. We adopted an analytical form of the intensity for a hotspot described by equation \ref{surface_profile} and generated $10^4$ sets of randomly distributed $\alpha$ (the angle between the rotation and magnetic axes) and $\zeta$ (the angle between the rotational axis and the line of sight). We assumed the Hopf beaming function \citep{Chandrasekhar1950} and calculated the pulse profiles (see Appendix \ref{modeling} for the detailed calculation). The gravitational light-bending effect was approximated using the analytical formula derived by \citet{Beloborodov2002}. We then calculated $A_2$ and PF from all the simulated profiles. The resulting probability density distributions are plotted in Fig.~\ref{a2p_pf_distribution}. $A_2$ is almost uniformly distributed between 0 and 0.7, while PF is strongly concentrated below $\sim$0.12.

We then followed the procedure described in \citep{DedeoPN2001} to tune the intensity ratios, $r$, between two hotspots from $1$ to $9$. For each set of $r$, we fix one hotspot with a profile described by equation \ref{surface_profile}, and set the intensity of the other hotspot by multiplying the same equation by $r$. Then we performed $10^4$ Monte Carlo simulations of different viewing geometries. We plotted three cases, $r=1$, $r=3$ and $r=9$, in Fig.~\ref{a2p_pf_distribution}. Not surprisingly, the distributions for $r=3$ and $r=9$ show low $A_2\lesssim0.2$, while that for $r=1$ shows a flat distribution because the asymmetric surface intensity profile tends to result in a single-peaked profile. PF shows a relatively broad distribution for $r=3$ and $r=9$, although profiles with high PF are slightly more probable in both cases. The sharp distribution for $r=1$ at $\rm{PF}\lesssim0.1$ indicates that symmetric antipodal hotspots never result in a pulse profile with a high PF. These three distributions are far from the observed one.We found that if we performed a simulation with a choice of randomly distributed $r$ between $1$ and $6$, and randomly distributed $\alpha$ and $\zeta$, the resulting distribution of $A_2$ and $PF$ can roughly reproduce the observed distribution. This implies that the surface temperature distribution of magnetars varies greatly from source to source.

Fig.~\ref{a2p_pf_allcase} shows PF versus $A_2$ for simulated profiles compared with observed ones. For the case of two symmetric hotspots, PF is systematically low and $A_2$ spreads over a large range. As $r$ increases, the maximum allowed $A_2$ decreases and PF increases. For an assumed beaming function, no system is allowed to lie above the upper envelope of the distribution.We also plotted the envelopes for the cases of isotropic emission without beaming, the beaming effect in \citet{AdelsbergL2006}, and the strongest beaming of $\cos^3\theta'$ (see Appendix \ref{modeling}). We tried different degrees of concentration of hotspots by changing the numerator in equation \ref{surface_profile} as $\cos^n \theta_m$ and tuning $n$ from $n=2$ (least concentrated, equivalent to a large hotspot) to $n=8$ (most concentrated, equivalent to a small hotspot). We found that more concentrated hotspots with weaker beaming show similar behaviours of a less concentrated hotspot with a stronger beaming function. Only $\sim50$ per cent of the sample is below the envelope of isotropic emission, indicating that atmospheric beaming is necessary. Most magnetars can be interpreted by the Hopf beaming function, except for 1E~2259. This object could have more concentrated hotspots or a beaming function stronger than the Hopf function. All magnetars are enclosed by the envelope of strong beaming functions.Moreover, a profile with both high $A_2$ and high PF needs two extremely small hotspots with strong atmospheric beaming. None of the magnetars in our sample shows these properties.

\begin{figure*}
\begin{center}
\begin{minipage}{0.49\linewidth}
\includegraphics[width=0.95\textwidth]{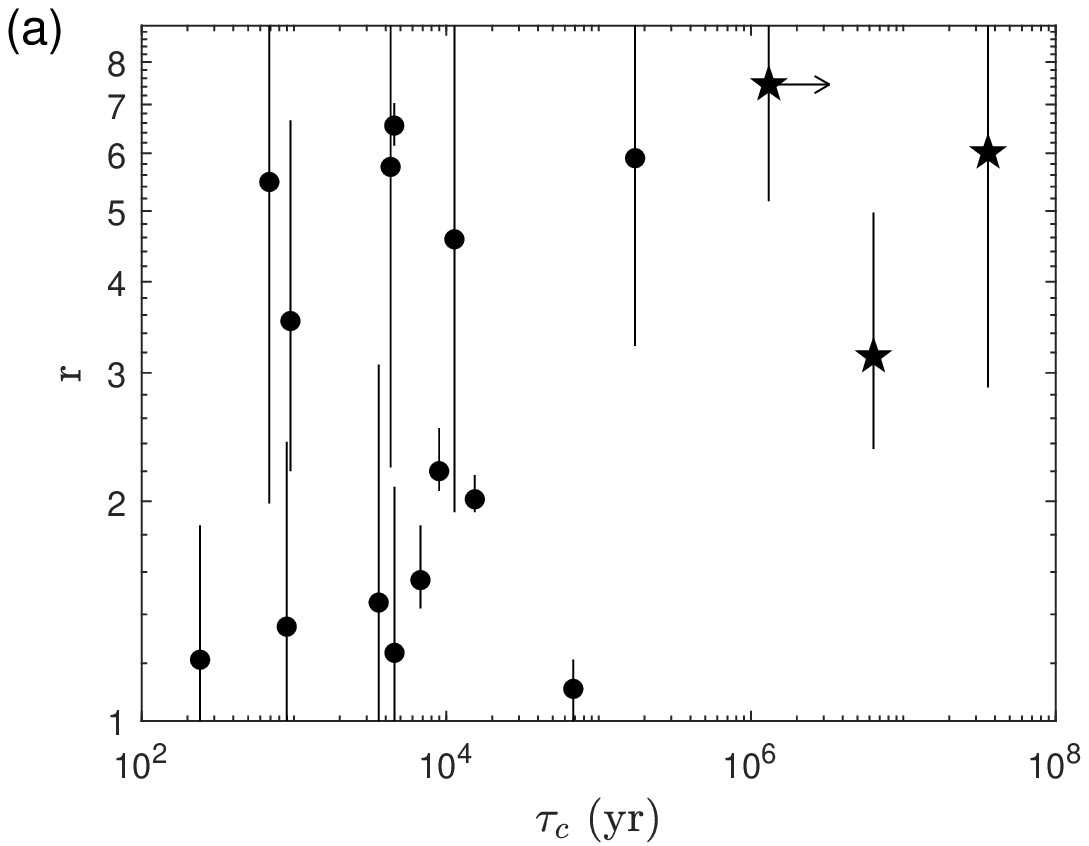}
\end{minipage}
\begin{minipage}{0.49\linewidth}
\includegraphics[width=0.95\textwidth]{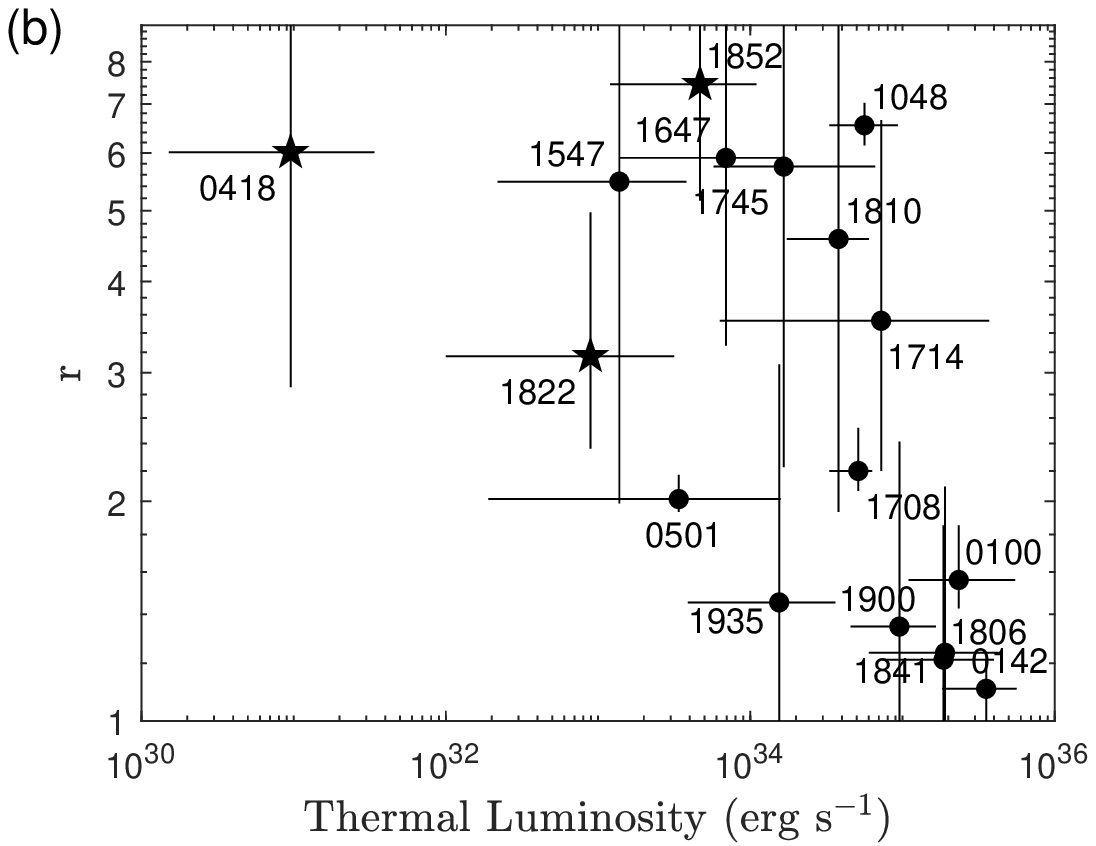}
\end{minipage}
\end{center}
\caption{Correlation between the inferred intensity ratio $r$ between two hotspots and (a) $\tau_c$ (b) the thermal luminosity. The star symbols refer to the low-$B$-field magnetars. The error bar of $r$ is calculated from the full range of the distribution in Fig.~\ref{magnetar_ratio}.  \label{lum_age_ratio}}
\end{figure*}

The above simulations suggest that the viewing geometry dominates the observed $A_2$ and PF. Hence, we estimated the degree of asymmetry for our sample of magnetars by obtaining the possible range of $r$ based on the simulation. We set up a $1000\times1000\times1000$ grid on $\alpha$, $\zeta$, $r$, and created simulated profiles by assuming a Hopf beaming. We obtained the distribution of allowed $\alpha$, $\zeta$, and $r$ that can produce the observed $A_2$ and PF of individual magnetars within the uncertainties. The distribution of $r$ is shown in Fig.~\ref{magnetar_ratio}.  Six magnetars, SGR~0418, 1E~1547, CXOU~J1647, SGR~1745, XTE~J1810, and 3XMM~J1852, have wide distributions truncated at the limit of $r=9$ in the simulation. Otherwise, their probability distribution of $r$ could extend to much higher values. We found that eight magnetars, CXOU~J0100, 4U~0142, SGR~0501, 1RXS~J1708, SGR~1806, 1E~1841, SGR~1900, and SGR~1935, have well-constrained $r\lesssim 3$. 1E~1048 has a well constrained $r$ between 6 and 7. CXOU~J1714 and Swift~J1822 have moderately constrained $r$ between 2 and 6. 1E~2259 is not included in Figure~\ref{magnetar_ratio} because the observed $A_2$ and PF (see Fig.~\ref{a2p_pf_allcase}) is beyond the limits of simulation based on our assumptions. However, a low value of $r$ is expected owing to its high $A_2$ and low PF.

We estimated the median value of the distribution for individual magnetars as the $r$ value and treated the boundaries of the distribution as the uncertainty intervals. The inferred values of $r$ are plotted against $\tau_c$ in Fig.~\ref{lum_age_ratio}(a). The plot shows no obvious correlation and no pattern similar to that in Fig.~\ref{age_parameter}. This suggests that the apparent evolutionary pattern in Fig.~\ref{age_parameter}(a) could be merely a coincidence. We also plotted the thermal luminosity dependence of $r$ in Fig.~\ref{lum_age_ratio}(b) and found a similar distribution to the PF–luminosity plot. The linear correlation coefficient between $r$ and luminosity is $-0.57$ with a null hypothesis probability of $0.009$, while Spearman’s rank correlation suggests a similar coefficient of $-0.64$ and a bit more significant probability of $0.006$. We cannot draw a strong conclusion for the linear correlation but a systematic trend probably exists. This implies that magnetars with high luminosities tend to have symmetric surface intensity profiles. It is necessary to include more magnetars to unambiguously confirm this correlation with the next generation of X-ray observatories.

\section{Discussion}\label{discussion}
Our results suggest a possible hint of anticorrelation between the intensity ratio of two antipodal hotspots and the thermal luminosity. This can be interpreted with the magneto-thermal evolution model. A magnetar is born as an extremely high-$B$-field NS and the $B$-field decay provides the energy to heat the surface \citep{KaminkerYP2006, PonsG2007, PonsMG2009, PernaP2011, ViganoRP2013, KaminkerKP2014}. The characteristic age $\tau_c$ may not be a good indicator for the evolutionary stage. Instead, the thermal luminosity could give us a handle on the magnetar age. Magnetic field evolution in NSs operates on time-scales of Ohmic diffusion,
\begin{equation}\label{ohmic}
\tau_{\rm Ohm}=4\pi\sigma_{\rm c}L^2/c^2 \sim 10^5\mbox{ yr}\left( \frac{\sigma_{\rm c}}{10^{23}\mbox{ s}^{-1}} \right) \left( \frac{L}{500\mbox{ m}} \right)^2,
\end{equation}
where $\sigma_{\rm c}$ is electrical conductivity and $L$ is the length-scale over which the magnetic field changes, and Hall drift,
\begin{equation}\label{hall}
\begin{aligned}
\tau_{\rm Hall} &= 4\pi en_{\rm e}L^2/cB \\
 &\sim 5\times 10^5\mbox{ yr} \left( \frac{\rho}{10^{13}\mbox{ g cm$^{-3}$}}\right) \left(\frac{B}{10^{14}\mbox{ G}} \right)^{-1} \left(\frac{L}{500\mbox{ m}}\right)^2, 
\end{aligned}
\end{equation}
where $n_{\rm e}$ is the electron number density and $\rho$ is the mass density \citep{GoldreichR1992, GlampedakisJS2011}. Hall drift plays an important role in destroying the surface temperature symmetrywhen the $B$ field is strong because it can generate small-scale structures and change the magnetic field geometry \citep{ViganoRP2013}. The degree of anisotropy depends on the strength of the initial toroidal field \citep{PernaVP2013}. If a neutron star only has a weak dipolar $B$ field, the magnetic poles will be hotter than the equatorial region, although the exact surface profile distribution depends on the location of the magnetic energy dissipation. The surface temperature distribution is symmetric with respect to the magnetic equator. If a neutron star has a strong toroidal field that contributes $\sim$90 per cent of the magnetic energy, an asymmetric temperature profile is expected as a result of the Hall drift \citep{GlampedakisJS2011}. In this case, the thermal pulse profile eventually evolves to single-peaked with a high PF. The observed correlation can be explained if all magnetars have strong toroidal fields. Luminous magnetars are young objects for which the surface temperature symmetry has not yet been destroyed. Of course, different initial temperature and $B$-field configurations could result in different evolutionary time-scales for the pulse profiles. These effects cause a large scatter in the $r$-luminosity plot (see Fig.~\ref{lum_age_ratio}). 

We found that those magnetars with intensity ratio $r\lesssim3$ are young objects, except for 4U~0142. 1E~2259 is also probably in this category because it shows a high $A_2$ and relatively low PF, although a stronger beaming function or a smaller hotspot is necessary. The SNR age of 1E~2259 is 14\,kyr, much younger than its $\tau_c=230$\,kyr, and this object can still be classified as a young magnetar. 4U~0142 could have a similar property, because its luminosity is extremely high, but further investigation of its true age is needed. In contrast, other magnetars may need high surface temperature asymmetry, although some of them have large error bars and $r<3$ remains possible. We are unable to draw strong
conclusions about their surface temperature anisotropy. All the low-$B$-field magnetars with large $\tau_c$ are classified in this category. They could be evolved magnetars in which the $B$ field decayed to the current low values and the symmetry has been broken. 1E~1048 is an intriguing case that the viewing geometry can be well constrained. It has a low $\tau_c=4.5$\,kyr and high spin-down inferred $B=3.9\times10^{14}$\,G, so is much younger and more
luminous than those low-$B$-field magnetars showing similar pulse profile properties. These behaviours could provide a hint of the initial $B$-field strength of magnetars. The Hall drift breaks the symmetry of the surface temperature distribution on a shorter timescale for a stronger $B$ field. As one of the eight extreme magnetars characterized by large $\dot{P}$ and high initial magnetic fields, 1E~1048 has the lowest thermal luminosity \citep{ViganoRP2013}. Therefore, it is probably an evolved extreme magnetar, and the symmetry was broken faster than in other magnetars with lower initial magnetic fields.

Alternatively, these outliers could be interpreted as arising from the effect of relatively small-scale hotspots on the NS surface. These hotspots can be generated by interior magnetic field structures composed of a mixture of dipolar and toroidal components, and could produce large PF, $A_2$, and/or hotspot asymmetry $r$. The generation and the persistence of these small-scale structures are attributed to Hall drift, which causes the movement of the $B$ field to regions of lower conductivity and subsequent enhancement of field dissipation and heating. These hotspots can grow quickly and persist for a long time, as demonstrated by numerical simulations of $B$-field evolution that are sometimes coupled to thermal evolution simulations. For example, \citet{GeppertV2014} found rapid development (on a timescale of $\sim 10^4$~yr) of substantial tens of degree surface regions with magnetic field and temperature exceeding global averages and these regions can last for $>10^6$\,yr. Similarly, \citet{GourgouliatosWH2016} and \citet{GourgouliatosH2018} found kilometre size magnetic spots whose field strengths can greatly exceed the dipolar $B$ field at the pole. However, it is important to keep in mind that the development of these magnetic field structures depends strongly on the initial magnetic field configuration, which is not uniquely prescribed, and requires a toroidal component at least as strong as the dipolar component \citep[c.f.~][]{KojimaK2012}.

\section{Summary}\label{summary}

We have carried out a comprehensive investigation of the quiescent soft X-ray pulse profiles of magnetars by calculating the strength of the Fourier components and PF. We find that over half of our sample of magnetars have low amplitudes of the second Fourier harmonic and low PF, while the others have either high $A_2$ or high PF. We further performed simulations to explore the surface temperature distribution by assuming two hotspots with different intensities. We find that the viewing geometry dominates the shape of the observed pulse profiles, and a diversity of the intensity ratio between two hotspots is needed to explain the observed distribution of $A_2$ and PF. The correlation analysis shows intriguing dependences between the profile shape and the physical parameters, including $\tau_c$, dipolar $B$ field, and the thermal luminosity. We estimated the intensity ratio between two hotspots by comparing the profile parameters from the magnetar sample and the simulations. We suggest that the surface temperature symmetry correlates with the thermal luminosity; that is, magnetars with higher luminosity generally have more symmetric profiles with respect to the magnetic equator. This can be interpreted as the result of evolution if all the magnetars have strong toroidal fields and the symmetry between two hotspots is destroyed through evolution. 

In addition to the main result, we updated the long-term spin period evolution of CXOU~J1714 with four more datasets after the latest reports. The result shows $\dot{P}=(6.41\pm0.03)\times10^{-11}$\,s\,s$^{-1}$ over a time span of $\sim$7 years, consistent with that obtained from the first two \chandra\ and \xmm\ observations. 

\section*{Acknowledgements}
We sincerely thank the referee for valuable suggestions on the paper. This research is in part based on the data obtained from the \emph{Chandra} Data Archive, and has made use of software provided by the Chandra X-ray Center (CXC) in the application packages CIAO, CHIPS, and SHERPA. This research has used observations obtained with \emph{XMM-Newton}, and the ESA science mission, with instruments and contributions directly funded by ESA Member States and NASA. C-PH and C-YN are supported by a GRF grant from the Hong Kong Government under HKU 17300215P. WCGH acknowledges support from STFC in the UK through Grant No. ST/M000931/1.





\bibliographystyle{mnras}
\input{magnetar_mnras.bbl}




\appendix

\section{Spin Period Evolution of CXOU J171405.7$-$381031}\label{new_pdot}
\begin{figure}
\centering
\includegraphics[width=0.5\textwidth]{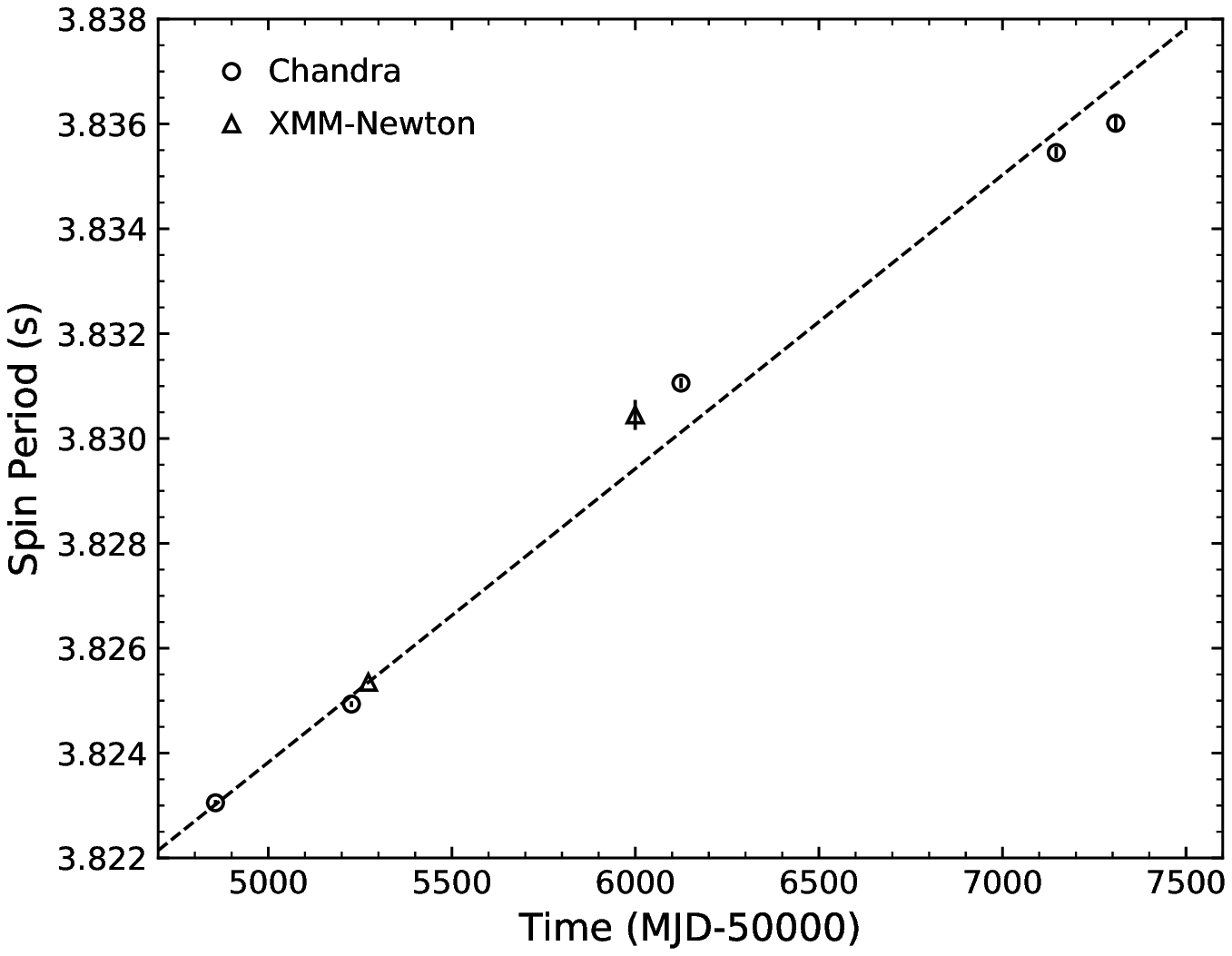}
\caption{Spin period evolution of CXOU J171405.7$-$381031 obtained with \chandra\ and \xmm.  \label{spin_evolution_cxouj171405}}
\end{figure}

Currently, the most reliable measurement of the spin-down rate of CXOU~J1714 is $\dot{P}=(6.40\pm0.05)\times10^{-11}$\,s\,s$^{-1}$ utilizing \chandra\ and \xmm\ observations in 2009--2010 (time span $\sim1.1$\,years). Because three more \chandra\ observations (ObsIDs \href{http://cda.harvard.edu/chaser/viewerContents.do?obsid=13749}{13749}, \href{http://cda.harvard.edu/chaser/viewerContents.do?obsid=16762}{16762}, \href{http://cda.harvard.edu/chaser/viewerContents.do?obsid=16763}{16763}) and one \xmm\ observation (ObsID \href{http://nxsa.esac.esa.int/nxsa-web/\#obsid=0670330101}{0670330101}) have been made since the latest reported $\dot{P}$, we performed the $H$-test to search for periodicity from all available \chandra\ and \xmm\ data to update the ephemeris. The evolution of the spin period is shown in Fig.~\ref{spin_evolution_cxouj171405}, where the linear fit implies a period derivative of $\dot{P}=(6.41\pm0.03)\times10^{-11}$\,s\,s$^{-1}$ over a time span of $\sim$7 years. This value is consistent with that determined by \citet{SatoBN2010}. Moreover, the fitting is poor, with a $\chi^2/dof=350/5$ and a significant discrepancy between individual data points and the long-term evolutionary trend is clearly seen. The difference is particularly large near $\sim$MJD 56000 where a timing anomaly event probably have occurred.

\section{Thermal Luminosities of Quiescent Magnetars}\label{spectral_analysis}
The emission of most magnetars shows two thermal components: a low-temperature component from the entire surface and a hightemperature component from the hotspots. The parameters of the low-temperature component are sometimes not well determined \citep[][ and references therein]{MongN2017}. We calculated the bolometric thermal luminosities from both components for those magnetars with well-constrained parameters from \citet{MongN2017}. Otherwise, we calculated the luminosities using a single blackbody (BB) model. We further fitted the X-ray spectra of SGR~0418, SGR~1745, 3XMM~J1852, and SGR~1935 extracted from the data sets listed in Table \ref{summary_dataset} to constrain their spectral parameters and the thermal luminosities in quiescence.  We used Sherpa \citep{FreemanDS2001} to fit the X-ray spectra. The absorption model we used in the spectral analysis was `tbnew\footnote{\url{http://pulsar.sternwarte.uni-erlangen.de/wilms/research/tbabs/}}'. We set the interstellar abundance according to \citet{WilmsAM2000}, which uses the cross section presented in \citet{VernerYB1993}. In case of multiple observations of a single source, we set the parameters to be the same between data sets because no significant variabilities were found. We fitted the \chandra\ spectra in 0.5--7\,keV and the \xmm\ spectra in 0.3--10\,keV with the Cash statistic \citep{Cash1979}.We used a simple BB to characterize the thermal component and added a power-law component when necessary. Adding another BB component is not necessary for these four sources. The results for SGR~0418, and SGR~1745, and 3XMM~J1852 are consistent with literature \citep{ReaIP2013, ZelatiRT2017, ZhouCL2014}. The spectral behaviour of SGR~1935 of the new \xmm\ dataset was not reported in the literature. We fitted the spectra and found that adding a power-law component can greatly improve the fit statistic. The maximum likelihood ratio test suggests a null hypothesis probability of $10^{-52}$. The temperature is consistent with previous \chandra\ observations \citep{IsraelER2016}. The best-fitting parameters, including the hydrogen column density $N_{\textrm{H}}$, the B temperature $kT_{\textrm{BB}}$ and normalization, and the photon index $\Gamma$, are shown in Table~\ref{spectral_fit_table}.  The thermal luminosities of all the magnetars were summarized in Table~\ref{summary_table}.

\begin{table*}
	\centering
	\caption{Best-fitting spectral parameters of SGR 0418, 3XMM~J1852, SGR~1745, and SGR~1935. }
	\label{spectral_fit_table}
	\begin{threeparttable}
	\begin{tabular}{lccccc} 
		\hline
Name & $N_{\rm{H}}$ ($10^{22}$\,cm$^{-2}$) & $kT_{\rm{BB}}$ (keV) & BB Norm$^a$ & $\Gamma$ & $C^2/dof$\\
\hline
SGR~0418 & $0.3\pm0.1$ & $0.33\pm0.03$ & $2.2_{-0.8}^{+1.0}\times10^{-4}$ & -- & 225.0/261 \\
SGR~1745 & $19_{-2}^{+3}$ & $0.67_{-0.04}^{+0.05}$ & $(9_{-4}^{+6})\times10^{-4}$ & -- & $703.1/744$ \\
3XMM~J1852 & $1.9\pm0.2$ & $0.69\pm0.02$ & $3.3_{-0.5}^{+0.7}\times10^{-4}$ & -- & $685.2/650$ \\
SGR~1935 & $2.3_{-0.2}^{+0.6}$ & $0.47_{-0.06}^{+0.03}$ & $(2.8_{-0.4}^{+0.3})\times10^{-3}$ & $1.8_{-0.6}^{+0.5}$& $438.7/452$ \\
\hline
	\end{tabular}
	\begin{tablenotes}
            \item[a] The normalization of the BB component is given by $9.884\times10^{31}\left( R_{\rm{BB}}/D \right)^2$, where $R_{\rm{BB}}$ and $D$ are the radius of and distance to the blackbody source in the same unit.
        \end{tablenotes}
        \end{threeparttable}
\end{table*}


\section{Modeling the Pulse Profile}\label{modeling}
\begin{figure}
\centering
\includegraphics[width=0.49\textwidth]{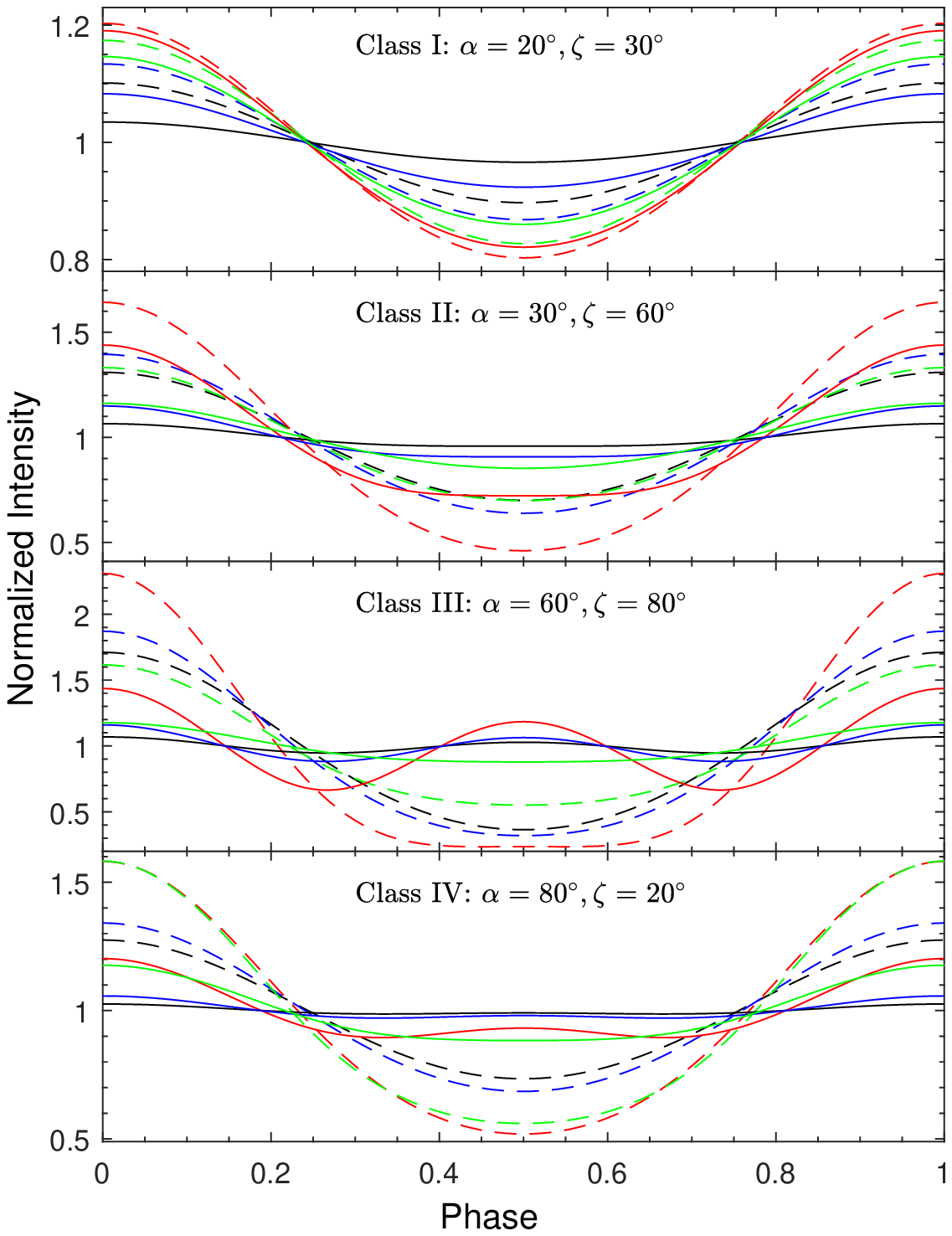}
\caption{Simulated pulse profiles for different $\alpha$ and $\zeta$. The solid profiles are from two hospots with equal brightness, while the dashed profiles are from two asymmetric hotspots in which one is nine times brighter than the other. The black, blue, red, and green curves are calculated from the isotropic emission, mild beaming from the Hopf function, strong beaming as $\cos^3\theta'$, and the beaming function adopted from \citet{AdelsbergL2006}. \label{simulate_profile}}
\end{figure}

To simulate the thermal pulse profile emitted from hotspots on a neutron star, we first adopted an analytic intensity profile presented in \citet{DedeoPN2001}:
\begin{equation}\label{surface_profile}
I_0(\theta_m,\phi_m)\sim I_0\frac{\cos^2\theta_m}{(3\cos^2\theta_m+1)^{0.8}},
\end{equation}
where $\theta_m$ and $\phi_m$ are the colatitude and the azimuth angle of the spherical coordinate with respect to the magnetic axis. A strong concentration of emission, which is equivalent to a smaller hotspot, can be achieved by increasing the power of the cosine function. This profile worked well for a dipolar $B$ field. The strong gravitational field near the neutron star bends the light emerging from the surface. To calculate the gravitational bending of the photon path, we used the approximate formula
\begin{equation}\label{deflaction}
1-\cos\theta'=(1-\cos\psi)\left( 1- \frac{r_s}{R} \right),
\end{equation}
where $\theta'$ is the angle between the emitted photon and the normal vector with respect to the stellar surface, $\psi$ is the photon escape direction observed from the infinity with respect to the normal vector, and $r_s=2GM/c^2$ is the Schwarzschild radius of an NS with a mass $M$ \citep{Beloborodov2002}. We set a typical NS radius $R=3r_s\approx10$\,km for which this approximation works well. For a given angle $\alpha$ between the rotation and magnetic pole axes, and an angle $\zeta$ between the rotational axis and the line of sight, the photon escape direction from a point on the surface can be written as 
\begin{equation}\label{rotation}
\cos\psi=\sin\alpha\sin\zeta\cos\omega t+\cos\alpha\cos\zeta,
\end{equation}
where $\omega$ is the angular frequency. With this approximation, the observed flux d$F$ from any surface element d$S$ can be written as
\begin{equation}\label{df}
\textrm{d} F=\left( 1-\frac{r_s}{R} \right)^2I_0(\theta,\phi,\theta')\cos\theta'\frac{\rm{d}S}{D},
\end{equation}
where $I_0(\theta,\phi, \theta')$ is the intensity profile in spherical coordinate $(\theta,\phi)$ with respect to the rotational axis, and $D$ is the distance between the observer and the NS \citep{DedeoPN2001, Beloborodov2002}. For isotropic emission, $I_0(\theta,\phi,\theta')$ is independent of $\theta'$. However, the emission from the NS surface is usually beamed owing to anisotropic scattering in the atmosphere and absorption of photons in magnetized plasmas. Fortunately, the gap between the narrow pencil beam and broad fan beam is reduced, which results in a featureless broad beam for highlymagnetized NSs \citep{AdelsbergL2006, PernaVP2013}. This will cause less complex pulse profiles in magnetars. We applied three simple beaming functions. The first one is the Hopf function with the fourth approximation:
\begin{equation}
I_0(\theta,\phi,\theta')=\frac{3}{4} I_0(\theta,\phi)\left(\sum_{a=1}^{3}\frac{L_a}{1+k_a\cos\theta'}+\cos\theta'+Q \right),
\end{equation}
where $k_a$ is the characteristic root, and $L_a$ and $Q$ are constants of integration. The values of $k_a$, $L_a$ and $Q$ are adopted from Table III.VII of \citet{Chandrasekhar1950}. This beaming function is suitable for a scattering dominated atmosphere. Another conventional beaming form is 
\begin{equation}
I_0(\theta,\phi,\theta')=I_0(\theta,\phi)\cos^n\theta',
\end{equation}
where a larger $n$ means a stronger beaming \citep{Nagel1981,DedeoPN2001}. This form well describe the beaming caused by the accretion, and we used $n=3$ to represent the heavily beamed case. Finally, we also adopted the intensity profile derived by \citet{AdelsbergL2006} that considered the vacuum polarization effects for the strong $B$ field of $B=5\times10^{14}$\,G.

We integrated equation (\ref{df}) over the visible surface with $\cos\psi>-r_s/(R-r_s)$ to obtain the pulse profile. Four typical profiles were shown in Fig.~\ref{simulate_profile}: class 
\uppercase\expandafter{\romannumeral1} ($\alpha=20$\,\degr, $\zeta=30$\,\degr), class 
\uppercase\expandafter{\romannumeral2} ($\alpha=30$\,\degr, $\zeta=60$\,\degr), class 
\uppercase\expandafter{\romannumeral3} ($\alpha=60$\,\degr, $\zeta=80$\,\degr), and class 
\uppercase\expandafter{\romannumeral4} ($\alpha=80$\,\degr, $\zeta=20$\,\degr) . For the case of symmetric hotspots, the pulse profile is single peaked for class \uppercase\expandafter{\romannumeral1} and class \uppercase\expandafter{\romannumeral2}, while it becomes double peaked for class \uppercase\expandafter{\romannumeral3} and class \uppercase\expandafter{\romannumeral4}.  Moreover, PF is enhanced as a result of the beaming effect. We also plotted the extreme case that one hotspot is nine times brighter than another. In this case, the profiles are all single-peaked with lower $A_2$ and higher PF compared to the symmetric case. 

To further test the available PF and $A_2$ from the asymmetric surface temperature profile, we tuned the different intensity ratio emerging from two hotspots from 1 to 9 and performed Monte Carlo simulations. For each assumed intensity ratio, we generated $10^4$ sets of randomly distributed $\alpha$ and $\zeta$, and calculated their pulse profiles, PFs, and $A_2$. The results are plotted in Fig.~\ref{a2p_pf_allcase}. The envelope denotes the available $A_2$ and PF with the brightness function described in equation (\ref{surface_profile}) with different beaming functions. We performed the same simulations for the case of more concentrated hotspots and found that the effect is similar to a stronger beaming.


\bsp	
\label{lastpage}
\end{document}